\documentclass[traditabstract]{aa}
\usepackage[varg]{txfonts}

\usepackage{graphicx}
\usepackage{natbib}
\usepackage{amsmath}
\usepackage{amssymb}
\usepackage{url}

\newcommand{\hde}[0]{HD~189733}
\newcommand{\hdz}[0]{HD~209458}
\newcommand{\lya}[0]{Lyman\,$\alpha$}

\newcommand{\ms}[0]{\mbox{m\,s$^{-1}$}}
\newcommand{\kms}[0]{\mbox{km\,s$^{-1}$}}
\newcommand{\rme}[0]{Rossiter-McLaughlin effect}
\newcommand{\rmc}[0]{Rossiter-McLaughlin}
\newcommand{\cahk}[0]{\ion{Ca}{ii} H and K}
\newcommand{\nal}[0]{\ion{Na}{i}}
\newcommand{\nadd}[0]{\ion{Na}{i}~D$_1$ and D$_2$}

\usepackage{marginnote}
\usepackage{color}
\begin{document}

\title{The center-to-limb variation across the Fraunhofer lines of
HD~189733\thanks{Based on observations made with UVES at the ESO VLT Kueyen
telescope under program 089.D-0701(A).}}
\subtitle{Sampling the stellar spectrum using a transiting planet}
\titlerunning{The CLV across the Fraunhofer lines of HD~189733}
\author{S. Czesla, T. Klocov\'a, S. Khalafinejad, U. Wolter, J.H.M.M.
Schmitt}
\institute{Hamburger Sternwarte, Universit\"at Hamburg, Gojenbergsweg 112, 21029 Hamburg, Germany}
\date{Received ... / Accepted ... }

\abstract{The center-to-limb variation (CLV) describes the brightness of the
stellar disk as a function of the limb angle. Across strong absorption lines,
the CLV can vary quite significantly. We obtained a densely sampled time series
of high-resolution transit spectra of the active planet host star HD~189733 with
UVES.
Using the passing planetary disk of the hot Jupiter \hde\,b as a probe, we study the CLV
in the wings of the \cahk\ and \nadd\ Fraunhofer lines, which are not strongly
affected by activity-induced variability. In agreement
with model predictions, our analysis shows that the wings of the studied
Fraunhofer lines are limb brightened with respect to the
\mbox{(quasi-)continuum}.
The strength of the CLV-induced effect can be on the same order as signals found
for hot Jupiter atmospheres. Therefore, a careful treatment of the wavelength
dependence of the stellar CLV in strong absorption lines is highly relevant in
the interpretation of planetary transit spectroscopy.
}

\maketitle

\section{Introduction}

 The discovery of the first \emph{\textup{transiting}} extrasolar planet \hdz\,b
\citep{Charbonneau2000} followed five years after the detection of the first extrasolar planet around a
main-sequence star \citep[51~Peg\,b,][]{Mayor1995}.
Both 51~Peg\,b and \hdz\,b are Jovian planets
orbiting their host stars at a distance of only a few hundredth of an AU, so-called hot Jupiters. At these distances, the atmospheres of the planets
are exposed to strong stellar radiation fields.
Unequaled in the solar system, the atmospheres of hot Jupiters can best be
studied in transiting exoplanetary systems, where they can be observed in
absorption against the bright background of the stellar disk.

In particular the stellar extreme ultraviolet and X-ray radiation
efficiently deposit energy in the upper planetary atmosphere, where it is
thought to derive a Parker-like planetary wind
\citep[e.g.,][]{Watson1981}. The wind material reaches beyond
the optical radius of the planet and effectively enlarges the radius of the disk,
especially at wavelengths were the absorption and scattering cross section of
the material is large, such as the hydrogen \lya\ line. Expanding planetary
atmospheres like this have, indeed, been detected by means of ultraviolet transit
spectroscopy, for instance, in the hot Jupiters \hde\,b and \hdz\,b
\citep{Vidal2003, Lecavelier2010}.
Further ultraviolet spectral features attributed to the upper planetary
atmosphere have been found in \hdz\,b and WASP-12\,b
\citep[e.g.,][]{Linsky2010, Fossati2010}. Even an X-ray detection of the
atmosphere of \hde\,b has been reported \citep{Poppenhaeger2013}.
 
The planetary atmosphere can also be probed by in-transit excess
absorption in the optical regime \citep[e.g.,][]{Seager2000, Brown2001}.
Variations in the planetary radius due to the presence of an atmosphere can be
probed, e.g., in the Fraunhofer \nadd\ lines at $\approx 5900$~\AA.
In 2002, \citeauthor{Charbonneau2002} announced the detection of
in-transit excess absorption at a level of
$(2.32\pm0.57)\times 10^{-4}$ in the \nadd\ lines of \hdz\
observed with the ``Space Telescope Imaging Spectrograph''
(STIS) on board the Hubble Space Telescope (HST).
While the first detection of atmospheric sodium was achieved with the HST, the
visual regime is also accessible with ground-based instrumentation.

After its space-based discovery,
the atmosphere of \hdz\,b was also the first to be detected with ground-based
instrumentation.
In a careful reanalysis of a data set earlier
discussed by \citet{Narita2005}, \citet{Snellen2008} found
excess
absorption in the \nadd\ lines of \hdz\,b at a
level of $(1.35 \pm 0.17)\times 10^{-3}$ in $0.75$~\AA\ wide spectral bands
centered on the \nadd\ line cores.
The signal is less pronounced when broader spectral bands are used.
Today, ground-based transmission spectroscopy is regularly used to
study planetary atmospheres;  a number of successful detections of
atmospheric sodium have been reported in the literature \citep[see,
e.g.,][]{Snellen2008, Wood2011, Sing2012, Zhou2012, Burton2015}.
% \LEt{According to A\&A style, the Introduction does not include any sections. Please either incorporate these into the Intro or into Section 2. }
% % \subsection{\hde}

The hot Jupiter \hde\,b is one of the best-studied planets beyond the solar
system to date. The planet orbits a K-type host star at a distance of $\approx 0.03$~AU
and transits the stellar disk every $\approx 2.2$~d \citep{Bouchy2005}. In
contrast to \hdz, \hde\ is a highly active star. The spectrum shows prominent
chromospheric cores in the \cahk\ lines and, consequently, a large S-index
of $0.525$ \citep[][]{Wright2004, Knutson2010}.
This is consistent with rotational photometric variability on the level of a few percent?
and an X-ray luminosity of $1.3\times 10^{28}$~erg\,s$^{-1}$, which makes \hde\
one of the most X-ray luminous K-type stars in the solar
neighborhood, and certainly puts it among the most active planet host stars
known to date \citep[][]{Schmitt1995, Croll2007, Poppenhaeger2013}.
Its distance of only $19.3$~pc and resulting apparent brightness of $V=7.67$~mag have
made it a prime target for detailed follow-up campaigns at all wavelengths
ranges. 

% \subsection{Transit spectroscopy of \hde\,b}

\hde\ was the target of an HST campaign carried out with
the Advanced Camera for Surveys (ACS) presented by \citet{Pont2008}. The
authors find an ``almost featureless'' spectrum between $5500$ and
$10\,500$~\AA. Later,
\citet{Huitson2012} presented another set of transmission spectra
of \hde\,b obtained with HST/STIS. In their analysis, \citet{Huitson2012} 
found excess absorption at a level of $(9\pm 1)\times 10^{-4}$ in the \nadd\
lines. The excess absorption is concentrated in the line cores, which
\citet{Huitson2012} attribute to planetary high-altitude haze or a
substantially subsolar sodium abundance of the planet. Both \citet{Pont2008} and
\citet{Huitson2012} found clear evidence for starspot occultations in their
light curves, which underlines the role of the high level of
stellar activity in \hde.

\citet{Redfield2008} reported on \nal\ excess absorption
in \hde, which is about twice as strong as that observed in \hdz\,b by
\citet{Snellen2008}.  \citet{Redfield2008} used a different convention
to quantify the excess absorption; a conversion has been carried out by
\citet{Snellen2008}. In a reanalysis of the data used by
\citeauthor{Redfield2008}, \citet{Jensen2011} confirmed their results and
reported excess absorption at a level of $(5.26\pm1.69)\times 10^{-4}$ across
a $12$~\AA\ wide spectral band.

Recently, \citet{Wyttenbach2015} reported on another
measurement of excess absorption at a level of $(3.2\pm0.3)\times 10^{-3}$ in
the \nadd\ lines of \hde\,b based on transit spectroscopy obtained with HARPS.
The quoted number refers to the same $0.75$~\AA\ wide bands used earlier by
\citet{Snellen2008}. \citet{Wyttenbach2015} conclude that the major part of
their signal is produced in the line cores and could even
quantify the width of the signal, giving a full width at half maximum
(FWHM) of $(0.52\pm0.08)$~\AA. Additionally, there seems to
be slight blue shift, which may be indicative of a wind in the planetary
atmosphere of \hde\,b blowing from the day to the night side, and is similar to what is observed in the case of \hdz\,b \citep{Snellen2010}.

While transit spectroscopy is a powerful technique for  studying the planetary
atmosphere, it is only seen in absorption against the stellar disk, whose
appearance is, therefore, equally relevant for an
appropriate interpretation of the results.
% \subsection{The center-to-limb variation}
\label{sec:CLV}
A wavelength-dependent change in the brightness of the
disk from the center toward the limb has long been observed on the Sun
\citep[e.g.,][]{Pierce1977, Neckel1994}.
The center-to-limb variation (CLV)
% \LEt{Please ensure that all abbreviations are
% written out at first mention, followed by the abbreviation in parentheses (even
% if you have already introduced them in the Abstract). After that please use
% only the abbreviation. Please check for your use of individual abbreviations
% throughout the paper. Instrument names do not need an introduction. .}
is
attributable to a limb-angle dependent viewing geometry, effectively exposing
ever higher layers of the solar atmosphere to the observer targeting the limb.
As higher layers are usually cooler and emit less light in the optical, the
effect is often more specifically
dubbed limb darkening. This term mostly, but not always, describes the
appearance of the stellar disk in the visual band.
At other wavelengths, such as in the X-ray and
ultraviolet regime, the solar atmosphere can show limb brightening
\citep[e.g.,][]{Schlawin2010, Poppenhaeger2013, Llama2015}.

In the analysis of transit light curves of extrasolar planets, the
CLV has long been acknowledged as a critical factor
\citep[e.g.,][]{Mueller2013}. In the modeling, the brightness distribution of
the stellar disk is usually parametrized by one of
many available so-called limb-darkening laws.
Among the most frequently applied parameterizations are the
linear and quadratic limb-darkening laws, according to which the limb-angle
dependent intensity is given by
\begin{equation}
  \frac{I_{\mathrm{linear}}(\mu)}{I(\mu=1)} = 1 - a\, (1-\mu)  
  \label{eq:lld}
\end{equation}
and
\begin{equation}
  \frac{I_{\mathrm{quadratic}}(\mu)}{I(\mu=1)} = 1 - a\, (1-\mu) - b\,(1-\mu)^2
  \; ,
  \label{eq:qld}
\end{equation}
where $a$ and $b$ are the
linear and quadratic limb-darkening coefficients,
and $\mu$ is the limb angle (see Sect.~\ref{sec:muDependSpecs}).
Limb-darkening coefficients for a wide range of
stellar parameters, limb-darkening laws, and photometric bands have been
tabulated \citep[e.g.,][]{Claret2004}.

While the tabulated coefficients usually refer to broad photometric bands,
the wavelength dependence of the CLV in the stellar atmosphere can be very
strong on shorter scales. In the optical, the wavelength dependence is strongest
across the profiles of deep stellar absorption lines, such as the \cahk\ or
\nadd\ lines \citep[e.g.,][]{Athay1972}.
\citet{Yan2015} presented a study on the CLV
in the solar Fraunhofer lines observed in solar eclipse spectra
reflected by the
Moon. The authors clearly detected the CLV of the solar
atmosphere in their spectra and argue that the effect  becomes most
prominent if narrow spectral bands covering the strong (Fraunhofer) lines are
used.

In fact, strong modulations in the stellar CLV over small wavelength ranges
also dubbed differential CLV (or differential limb darkening) are a known
problem in the analysis of the \nal\ features in extrasolar planets. For
instance, \citet{Charbonneau2002} provide a careful discussion on the effect to
rule out that their signal is caused by the CLV. 
Also \citet{Sing2008LD}  discuss the
CLV in \hdz\  extensively and present CLV-corrected transit light curves.
Among others, the authors conclude that limb darkening in the stellar lines is
smaller than in the continuum and that one-dimensional stellar atmospheres can
reproduce the observed effects.
In fact, many authors, including some
of those referenced earlier, have discussed the issue so that the above list
is by no means complete.

The treatment of the CLV is further complicated by the availability (or lack
thereof) of different model predictions. \citet{Claret2004} give both
limb-darkening coefficients derived from (one-dimensional) PHOENIX and ATLAS
model atmospheres, which are similar but not identical. \citet{Mueller2013}
find both sets of coefficients to be consistent with \textit{Kepler} data. 
\citet{Hayek2012} compare CLV predictions resulting from one- and
three-dimensional model atmospheres to transit light curves of \hde\ and \hdz.
For the latter, they find a considerably better match with the three-dimensional
model. While this is also compatible with their finding for \hde, their data
remained insufficient for a conclusive result in this case. 

In this work, we present the analysis of the CLV in the \cahk\ and
\nadd\ lines in \hde\ and discuss its relevance
for searches of atmospheric excess absorption in the \nal\ lines via transit
spectroscopy. In Sects.~\ref{sec:simul} to \ref{sec:DLCsstellarDisk}, we
introduce the formalism and models used in our analysis. Our observations and
the data analysis are presented in Sects.~\ref{sec:Observations} to
\ref{sec:CLV_Nadd}. Finally, we discuss the results in Sect.~\ref{sec:SummAndDis}
and present our conclusions in Sect.~\ref{sec:Conclusion}. 

\section{Synthetic data}
\label{sec:simul}

Throughout this work, we present synthetic spectra and simulated light curves 
of transiting planetary systems. Here, we outline
the simulations and  input data.

\subsection{Limb-angle dependent spectra}
\label{sec:muDependSpecs}
The limb angle, $\theta$, is the angle between the outward normal of the
stellar atmosphere and the vector pointing from the center of the star toward
the observer. Usually its cosine $\mu=\cos(\theta)$ is used to specify the
limb angle. With this convention, $\mu=0$ refers to the limb and $\mu=1$ to the
center of the stellar disk.

All synthetic spectra  used in this work are based on
Kurucz model atmospheres with solar metallicity \citep{Castelli2004, Kurucz1970}.
The spectra themselves were generated using the \texttt{spectrum}
program\footnote{http://www.appstate.edu/\textasciitilde
grayro/spectrum/spectrum.html} by R.O. Gray \citep{Gray1994},
which facilitates the generation of both
disk-integrated stellar spectra and limb-angle dependent specific intensities.
Based on the assumption of plane-parallel atmospheres and
local thermodynamic equilibrium (LTE), \texttt{spectrum}
first computes the number densities of
electrons and all relevant species in the atmosphere and solves the equation of
radiative transfer to obtain a synthetic spectrum. Information on individual
spectral lines, viz., their wavelength, the ion, the lower and upper energy
levels, and the product of statistical weight and oscillator strength are
specified via a line list file. In our calculations, we relied on the line list
shipped along with the program.

Whenever we use limb-angle dependent specific intensities, we generate
$20$~spectra with limb angles between $\mu=0$ and $\mu=1$, i.e., a spacing of
$\Delta\mu = 0.05$. To avoid numerical problems at the very edge of the stellar
disk ($\mu=0$), we replace the lower limit by $\mu=0.001$ in the calculations
and assume that the spectrum completely vanishes for $\mu=0$. The impact of
this approximation remains small because the stellar atmosphere is seen at a
very steep angle so close to the limb and only contributes a tiny fraction to
the total observed flux. Spectra at intermediate limb angles are linearly
interpolated based on the spectra on the grid. All spectra are
equidistantly sampled on an equidistant wavelength grid with a spacing of
$0.01$~\AA\ per bin.

\subsection{Discretized stellar surface}
To simulate the spectra of a stellar disk, potentially partially occulted by a
planet, we use a discretized stellar surface with elements defined
according to the prescription by \citet{Vogt1987}. In particular, we use $1001$
latitudinal rings and a total of $250\,000$ surface elements. In the
\citeauthor{Vogt1987} discretization, the surface elements have approximately
the same size, which is about $5\times 10^{-5}$~rad or $0.17$~square degree in
our case.

The spectrum is obtained by summing the contributions of all visible surface
elements, assigning to each a spectrum according to its projected area, limb
angle, and radial velocity. A transiting planet is implemented by occulting
all stellar surface elements covered by the planetary disk. This is equivalent
to setting their projected area to zero during occultation.  

Given this configuration, the surface element with the largest projected area
covers the $\pi \times (5\times 10^{-5})^{-1}=62\,500$th part of the
visible stellar disk. Therefore, a Jovian planet occulting one percent of the
disk would be sampled by about $625$ stellar surface elements, which gives an
idea of the statistical uncertainty in the simulations.

\section{Transit light curves and their combination}

Transit light curves, whether derived from broadband photometry or
high-resolution spectroscopy, are among the most important tools to study
exoplanets. Below, we briefly discuss the most relevant aspects
required in our work.

\subsection{The standard planetary system}
\label{sec:sp}
We  present a number of model calculations referring
to what we call our standard planetary system, which are used whenever
we do not discuss a specific planetary system such as \hde. The standard planet
represents a typical hot Jupiter. It has one Jovian radius and orbits a
star with solar radius in a distance of ten solar radii at an inclination of
$90$~degrees. The orbital period is $2$~days. The standard star is a slow
rotator with $v\sin(i)=0.1$~km\,s$^{-1}$ seen at an inclination of
$90$~degrees (i.e., equator-on).
Assuming the convention that the primary transit occurs at orbital phase zero,
the transit lasts from orbital phase $-0.018$ to $+0.018$, and the
transit duration is $1.69$~h. 

\subsection{Naming conventions for light curves}
\label{sec:naming}

We obtain the time-dependent flux, $f_f(t_i)$, in a spectral band centered
on some feature of interest and its equivalent, $f_r(t_i)$, in the reference
band, where the $t_i$ refer to the times of our hypothetical measurements. 
Further, we let $t_{\mathrm{it}}$ and $t_{\mathrm{oot}}$ refer to all
measurements in- and out-of-transit.
We dub $f_f(t_i)$ and $f_r(t_i)$ two light curves. In transit analyses,
these are often normalized by their out-of-transit levels, so that
\begin{equation}
  n_X(t_i) = \frac{f_X(t_i)}{\overline{f_X(t_{\mathrm{oot}})}} \; ,
\end{equation}
where $X$ represents $f$ or $r$ and the bar indicates the mean value. We 
refer to $n_X$ as the normalized light curve.
The difference, $d(t_i)$, of the normalized light curves, i.e.,
\begin{equation}
  d(t_i) = \frac{f_f(t_i)}{\overline{f_f(t_{\mathrm{oot}})}} -
  \frac{f_r(t_i)}{\overline{f_r(t_{\mathrm{oot}})}} = n_f(t_i) - n_r(t_i) \; ,
\end{equation}
traces differential changes in between them.
We dub $d(t_i)$ the difference curve (DC); this form was used by
\citet{Charbonneau2002} in their analysis. To quantify the difference in transit
depth, we follow \citet{Charbonneau2002} in defining the in-transit DC
excess (DCE)
% \LEt{If you are quoting Charbonneau et al., then keep the quotation marks. If not, the quotation marks are not necessary.}
\begin{equation}
  DCE = \overline{DC(t_{\mathrm{it}})} - \overline{DC(t_{\mathrm{oot}})} \; .
\end{equation}
We note that another natural choice for studying differential changes is the ratio,
$r(t_i)$, of the normalized light curves
\begin{equation}
  r(t_i) = \frac{f_f(t_i)}{f_r(t_i)} \times
  \frac{\overline{f_r(t_{\mathrm{oot}})}}{\overline{f_f(t_{\mathrm{oot}})}} =
  \frac{n_f(t_i)}{n_r(t_i)} - 1 \; ,
\end{equation}
where we subtract one to obtain the same out-of-transit level as for the DC.
Yet, the difference between the DC and the ratio curve
\begin{equation}
  DC(t) - r(t) = DC(t)\left( 1-\frac{1}{f_r(t)}\right) \; 
\end{equation}
is small
for typical planetary transit light curves with transit depths of a few percent.
Therefore, we limit the following discussion to the DC.

\subsection{The appearance of the difference curve}
\label{sec:diffandrat}

Assume we are given the normalized light curves of a feature, $LC_f(t_i)$, and
a number of normalized reference light curves, $LC_{r,1\ldots n}(t_i)$.
For instance, these light curves could have been extracted from a series of
spectra by integrating the signal in various bands.
In this particular case, we assume that $LC_f(t_i)$ and $LC_{r,1\ldots n}(t_i)$
are normalized transit light curves of our ``standard planetary system'' (see
Sect.~\ref{sec:sp}), which we obtain from conventional transit models
\citep{Mandel2002}. The only difference between the feature and reference light
curves is the stellar limb darkening. For the sake of simplicity and
without loss of generality, we assume the linear
limb-darkening law in this demonstration (see Eq.~\ref{eq:lld}).

\begin{figure}
  \includegraphics[angle=0, width=0.49\textwidth]{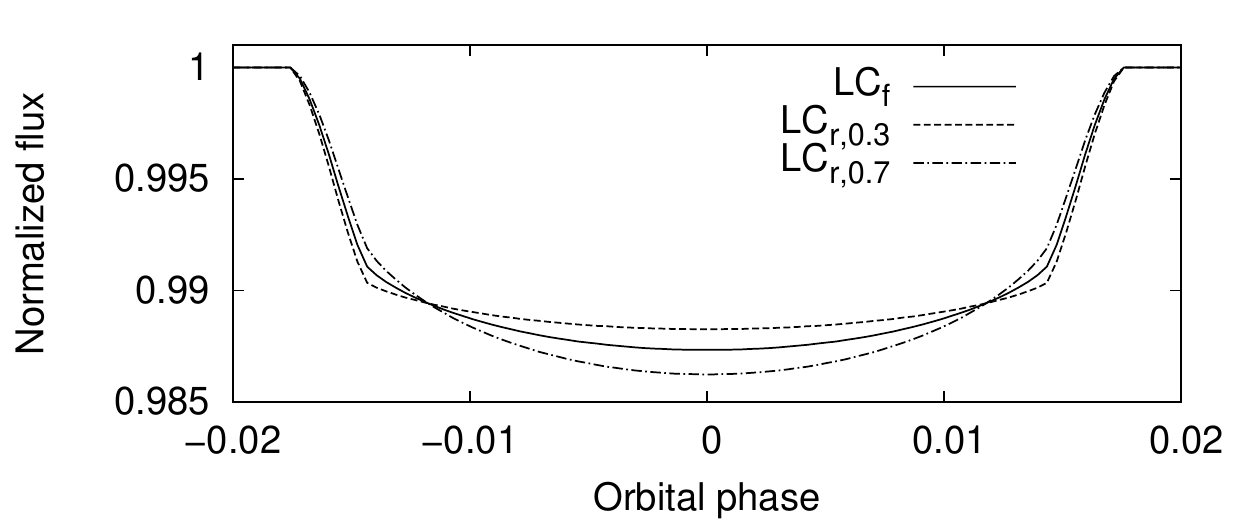}
  \caption{Normalized model transit light curves.
  \label{fig:refLCs}}
\end{figure}

For the feature light curve, $LC_f(t_i)$,
we specifically assume a value of $0.5$ for the linear limb-darkening
coefficient, $a$.
Then, we generate two reference light curves: the first, $LC_{r,0.3}(t_i)$,
using a limb-darkening coefficient of $0.3$ and another one, $LC_{r,0.7}(t_i)$,
assuming a value of $0.7$. The resulting model light curves are shown in
Fig.~\ref{fig:refLCs}.
Compared to $LC_{r,0.3}(t_i)$, the feature
shows limb darkening, but it shows limb brightening with respect to
$LC_{r,0.7}(t_i)$. We emphasize that all of the light curves show limb
darkening, and limb brightening is only seen as a relative effect.  

\begin{figure}[h]
  \includegraphics[angle=0, width=0.49\textwidth]{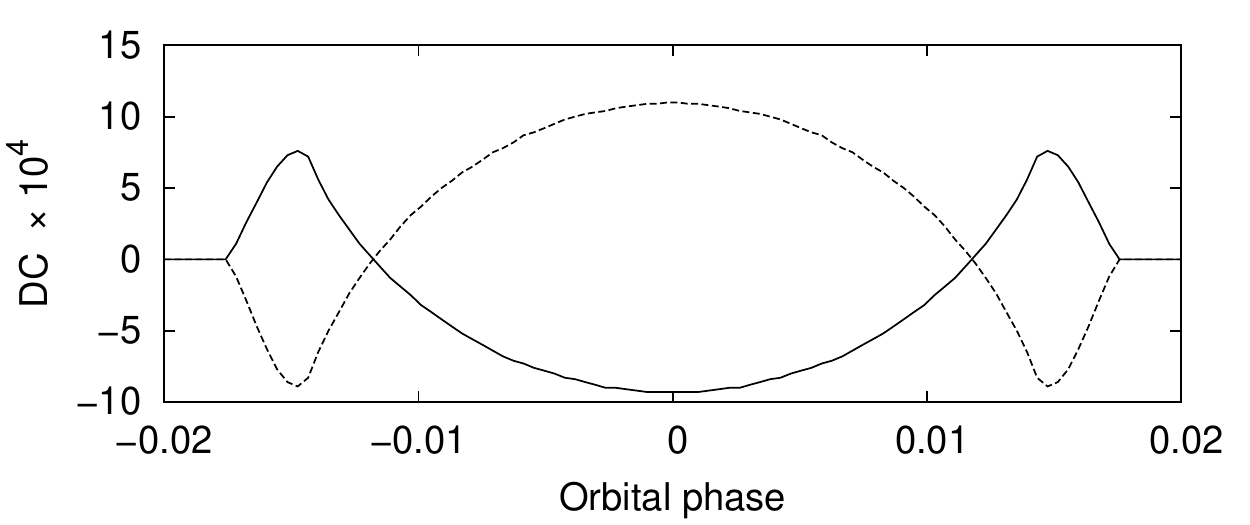}
  \caption{Difference curve, $DC(t_i)$,
  for a limb-darkened (solid, $LC_f(t_i)$ vs. $LC_{r,0.3}(t_i)$) and
  a limb-brightened (dashed, $LC_f(t_i)$ vs. $LC_{r,0.7}(t_i)$) feature.
  \label{fig:ldlb}}
\end{figure}

In Fig.~\ref{fig:ldlb}, we show both the DCs
obtained by combining the feature light
curve, $LC_f(t_i)$, with either of the reference light curves. The DCs 
show a distinct behavior, which differs substantially from that of the
input transit light curves (cf. Fig.~\ref{fig:refLCs}). In the
case of identical limb darkening for the feature and reference light curve, the
DC becomes zero.

When the feature is limb brightened compared to the reference light curve,
the DC shows a drop below the out-of-transit level during in- and egress.
At the limb of the stellar disk, the planet blocks
more light from the feature than it blocks from the reference. 
In the middle of the transit, with the planet  in the center of the stellar
disk, the DC shows a peak because the situation is reversed.
Compared to the feature, more of the reference light is emitted in the center of
the stellar disk, where it is now occulted by the planetary disk.
When the feature is limb darkened, the DC is inverted, but otherwise keeps its
main characteristics.

\section{Differential light curves and the stellar disk}
\label{sec:DLCsstellarDisk}

The presence of excess absorption attributable to an exoplanetary atmosphere
during transit is often derived by analyzing differential light curves, such as
the DC introduced in
Sect.~\ref{sec:naming}.
Usually, the spectral flux is integrated in a band
centered on the specific feature under consideration and compared to the
flux observed in various reference bands \citep[e.g.,][]{Charbonneau2002,
Snellen2008}. Depending on the characteristics of these bands, the
limb-angle dependence of the stellar spectrum, i.e., the stellar atmosphere,
can imprint a strong signal on the DC totally unrelated to the planetary
atmosphere.

\subsection{Limb-angle dependence of spectral lines}
\label{sec:ldspeclines}

\begin{figure}[h]
  \includegraphics[angle=0, width=0.49\textwidth]{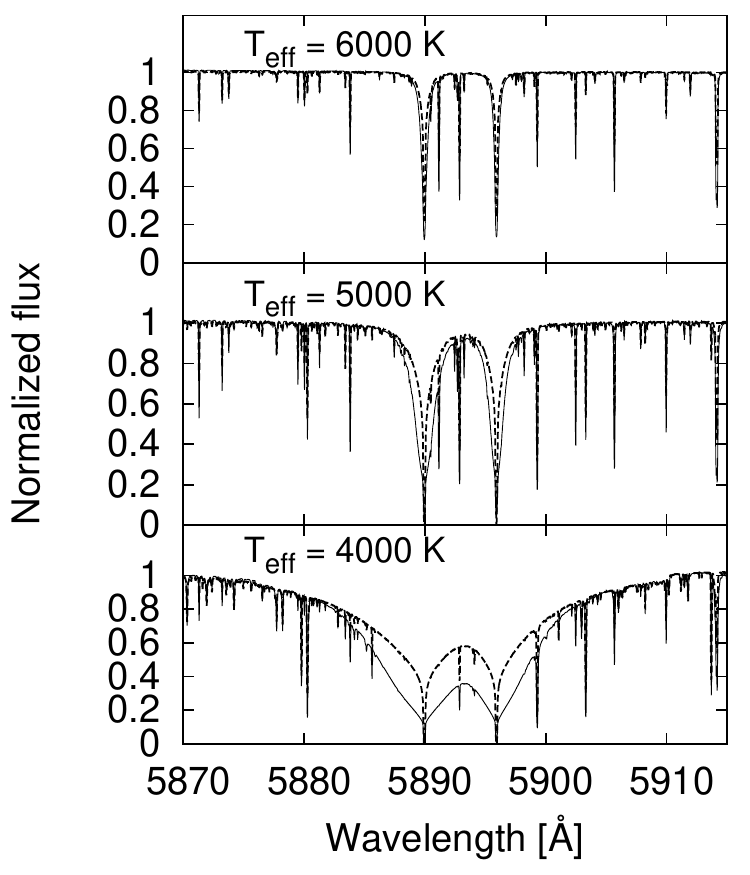}
  \caption{Normalized stellar spectrum around the sodium doublet for two limb
  angles $\mu = 0.1$ (dashed) and $\mu=1.0$ (solid) and three effective
  temperatures.
  \label{fig:Na}}
\end{figure}

In Fig.~\ref{fig:Na}, we demonstrate the effective temperature and limb-angle
dependence of the \nadd\ lines using stars with effective
temperatures of $4000$~K, $5000$~K, and $6000$~K.
Irrespective of limb angle, the width of the
\nadd\ lines increases when the effective temperature decreases.
This behavior can primarily be attributed to pressure broadening, viz., to
the temperature and pressure dependence of the line damping constants. In
solar-like stars, pressure broadening of the \nadd\ lines is dominated by
van der Waals forces \citep[][Fig.~11.4]{Gray2008}, whose influence is stronger
in the denser atmospheres of cooler stars.

At all temperatures, there is a strong dependence of the line
profile on the limb angle. In the center of the stellar disk, the line wings are more
pronounced than at the limb. At the stellar limb, higher layers with less
pressure are seen, where pressure broadening is weaker.
Although the limb-angle dependence is more clearly seen
in cooler stars, this is mainly attributable to the larger line
width. If the wavelength was scaled by the \nal\ line width,
the behavior of the line wings would actually become quite similar.
In contrast to the wings,
the depth of the line core remains virtually unchanged, which
has an immediate and important consequence: compared to the
core and the continuum, the line wings of the \nal\ doublet show
limb brightening. In fact, not only the \nal\ lines, but all spectral
lines show distinct CLV.

\begin{figure}[h]
  \includegraphics[angle=0, width=0.49\textwidth]{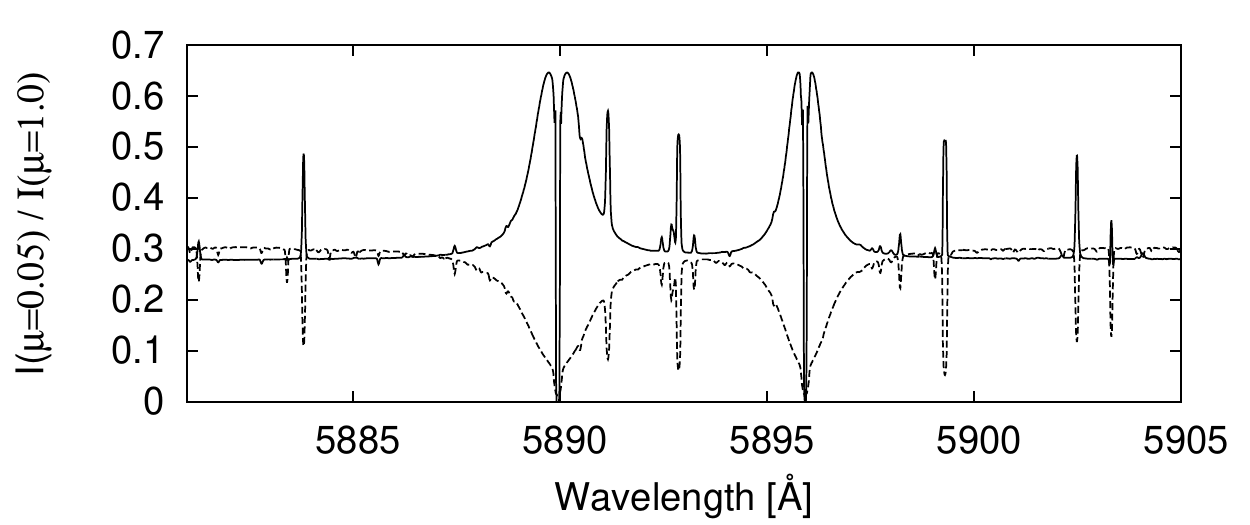}
  \includegraphics[angle=0,
  width=0.49\textwidth]{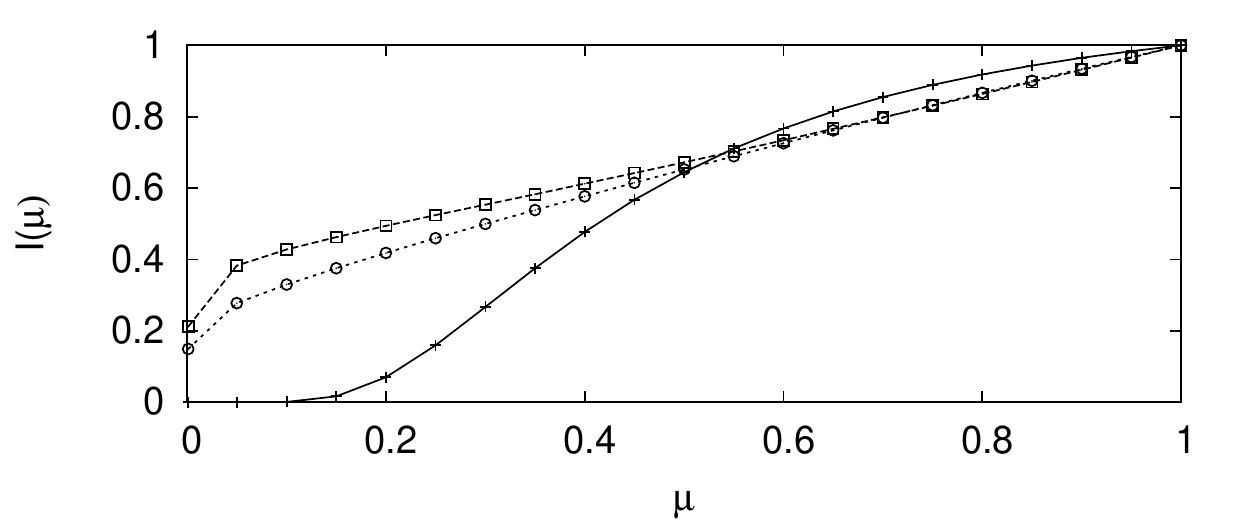}
  \caption{Top: the ratio of stellar spectra ($T_{\mathrm{eff}} = 5000$~K,
  $\log(g)=4.5$) at limb and center ($\mu=0.05$ and $\mu=1.0$) shown with a
  solid line. For comparison, the scaled central spectrum (dashed line). Bottom:
  normalized specific intensity in the $\mu=0.001-1$ range in the line core
  ($5889.95$~\AA, crosses), the line wing ($5889$~\AA, squares), and the quasi
-continuum ($5864.6$~\AA, circles).
  \label{fig:narat}}
\end{figure}

In Fig.~\ref{fig:narat}, we show the ratio, $R(\lambda)$, of the specific
intensity in the center ($\mu=1$) and at the limb ($\mu=0.05$) of a stellar disk
with an effective temperature of $5000$~K and $\log(g)=4.5$. In terms of
limb darkening (see Sect.~\ref{sec:CLV}), this ratio corresponds to
\begin{equation}
  R(\lambda) = \frac{I(0.05)}{I(1.0)} \; .
\end{equation}
While the (quasi-)continuum sections of the spectrum show a mostly constant
ratio, the spectral lines show distinct behavior.
Ignoring the sodium doublet for the moment, we find that $R(\lambda)$ increases
for the majority of spectral lines covered in Fig.~\ref{fig:narat}.
These lines, therefore, show limb brightening with respect to the
continuum, however, a number of lines also show the reverse effect.

Whether a spectral line shows limb brightening or darkening with respect to the
continuum depends on its height of formation in the stellar atmosphere. Lines
formed in the upper layers become stronger on the limb, where their depth and
equivalent width increases because the line of sight resulting from the
steeper viewing angle favors higher layers.
Spectral lines formed in deeper layers become stronger in the center of the
disk, where we look deeper into the atmosphere.
While the distribution of
spectral line formation in the stellar atmosphere depends on the details of the
stratification,
it can be said that the height of formation for lines originating from a single
ion tends to be higher for lower excitation potentials
\citep[e.g.,][]{Grossmann-Doerth1994}.

The strongest lines, such as the Na~I doublet, are formed over a wide range of
depths in the atmosphere. Therefore, they show the most complex behavior of the
spectral lines visible in Fig.~\ref{fig:narat} (top).
The line cores show immense optical thickness and are, therefore,
formed high up in the atmosphere. In fact, at $\mu=0.05$ the intensity in the
line cores has almost dropped to zero already (Fig.~\ref{fig:narat}, bottom).
This corresponds to an effective decrease in the stellar radius
or an increase in the planet-to-star radius ratio in the cores of the \nal\
lines.
While the
cores, therefore, show strong limb-darkening with respect to the continuum
emission, the line wings exhibit the opposite behavior. The wings are
limb brightened with the details depending on the exact
spectral band chosen.
Again, we emphasize that the line wings show only \emph{\textup{relative}}
limb brightening compared to the continuum. In the bottom panel
of Fig.~\ref{fig:narat}, we show the normalized specific intensity of the line
core, a point in the line wing, and a point in the surrounding quasi-continuum.
The curves clearly demonstrate that the line core shows limb darkening and the
line wing shows limb brightening relative to the continuum.

Figure~\ref{fig:narat} clearly
shows that the assumption of a homogeneously limb-darkened stellar disk becomes
inappropriate as soon as individual spectral lines are considered. 
The limb-angle dependence is most prominent
in deep and broad lines, such as the \cahk\ lines or the \nal\ doublet
lines. The latter are a main focus of the study
of exoplanet atmospheres \citep[e.g.,][]{Redfield2008, Charbonneau2002,
Snellen2008, Huitson2012}.

\subsection{Simulated Na~I transit light curves}
Using the techniques outlined in Sect.~\ref{sec:simul}, we simulated
spectral time series covering a transit of our standard exoplanetary system
(see Sect.~\ref{sec:sp}). In the simulations, we used a grid of stellar
effective temperatures ranging from $4000$~K to $7000$~K in steps of $500$~K;
$\log(g)$ was $4.5$ in all simulations.

We simulated the spectra for $60$
equidistant time points between orbital phases of $-0.02$ and $0.02$, where the
transit center occurs at phase zero. The temporal cadence of the simulated
spectra, therefore, is $(0.04\times 2~\mbox{d})/60 = 1.92$~min.
The numbers quoted in the following refer to spectra not broadened by
instrumental resolution, however, we found its impact  remains weak as long
the considered spectral bands cover a few spectral resolution elements.

\subsubsection{Difference curves with broad feature intervals}
In their analysis of the atmosphere of \hdz\,b, \citet{Charbonneau2002} used
bands covering the entire sodium-line doublet. In particular, they defined
narrow (n), medium (m), and wide (w) bands for the blue and red reference bands
and the central band covering the sodium lines.

\citet{Charbonneau2002} detect excess absorption by the planetary atmosphere
through a negative excess in what we dubbed the DC (see
Sect.~\ref{sec:naming}). As the authors point out, the CLV
is a potential source of this kind of an excess. However, they
verified that it is of no concern in their analysis because the effect remains
small (see their Eq.~6).

\begin{table}[h]
\centering
\caption{Simulated in-transit excess absorption diagnosed by the DC for the
narrow (n), medium(m), and wide (w)
spectral bands defined by \citet{Charbonneau2002}.
\label{tab:charbo}}
\begin{tabular}{l l l l}
\hline\hline
T$_{\mathrm{eff}}$ & n & m & w \\
$[$K$]$ & [$10^{-5}$] & [$10^{-5}$] & [$10^{-5}$] \\ \hline
4000 & 49.23 &  6.10 &  2.49 \\
4500 & 29.73 &  4.34 &  2.51 \\
5000 &  6.11 &  1.08 &  1.02 \\
5500 &  2.86 &  0.55 &  0.56 \\
6000 &  1.56 &  0.29 &  0.35 \\
6500 &  0.88 &  0.14 &  0.23 \\
7000 &  0.57 &  0.08 &  0.15 \\  \hline
\end{tabular}
\end{table}

We used the spectral bands defined by \citet{Charbonneau2002} and reproduced
their DCE estimates (see Sect.~\ref{sec:naming}) for our
simulations of the standard planetary system.
Our results are summarized in Table~\ref{tab:charbo}. Clearly, there is an effect
related to the CLV, which is stronger for lower effective temperatures
for the specified bands. We attribute this trend to the increasing width of the
Fraunhofer \nal\ lines, when effective temperature decreases.

\citet{Charbonneau2002} carried out the analysis for
\hdz\,b, which orbits a host star with an effective temperature of $6065$~K
\citep{Torres2008}. The values given by \citet{Charbonneau2002} read
$1.52\times 10^{-5}$, $0.39\times 10^{-5}$, and $0.47\times 10^{-5}$ for the
narrow, medium, and wide bands. These values are compatible with the values we
derived for a host star with an effective temperature of $6000$~K. We attribute
the small deviation to the difference in the system geometry and the different
model spectra.

\subsubsection{Difference curves with narrow feature intervals}
\begin{figure*}[t!]
  \includegraphics[angle=0, width=0.49\textwidth]{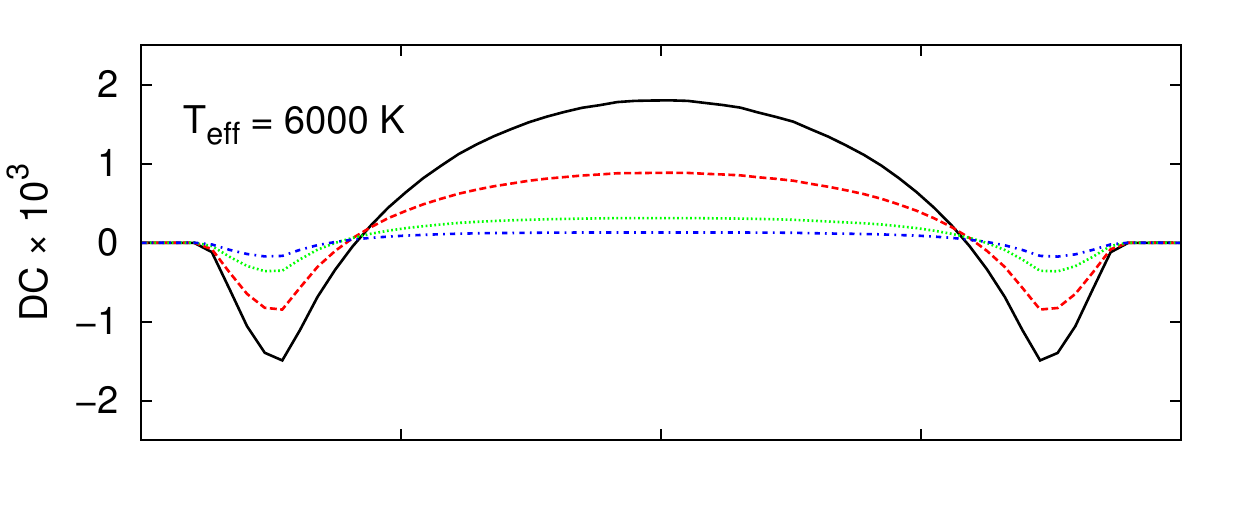}
  \includegraphics[angle=0, width=0.49\textwidth]{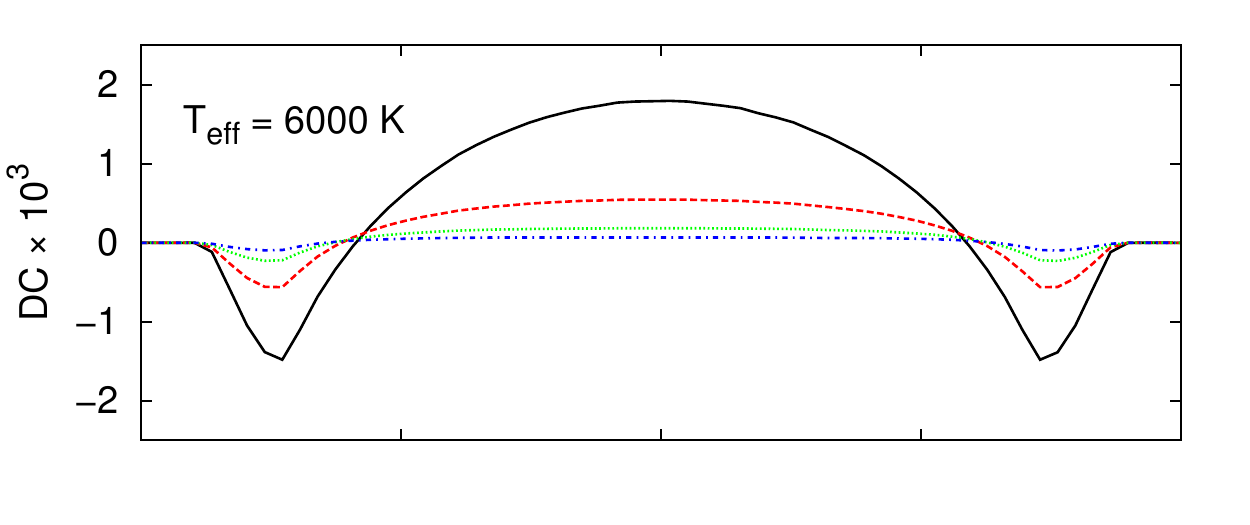}
  \\
  \includegraphics[angle=0, width=0.49\textwidth]{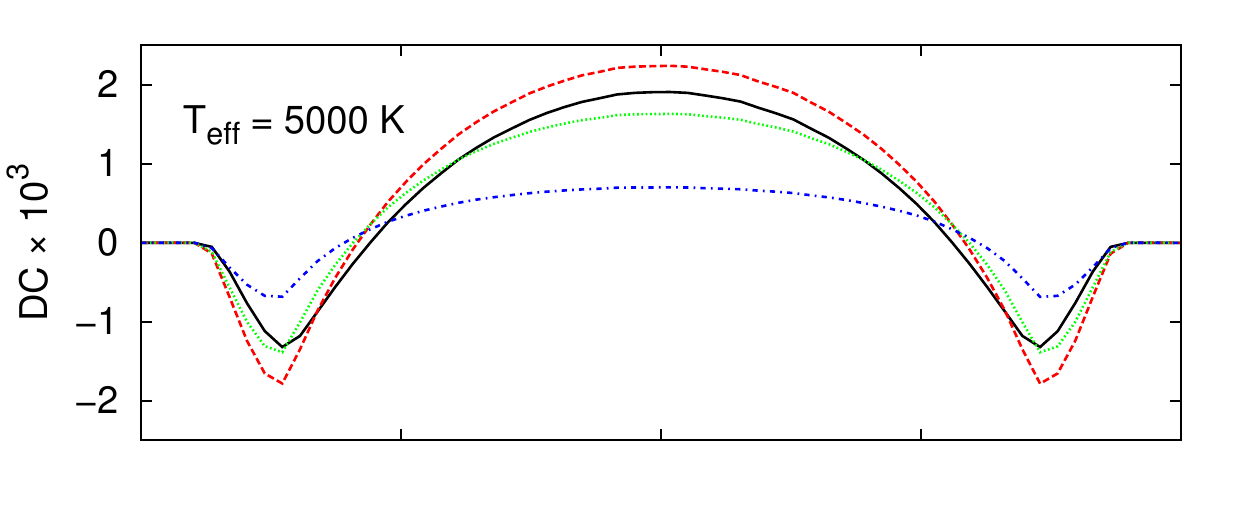}
  \includegraphics[angle=0, width=0.49\textwidth]{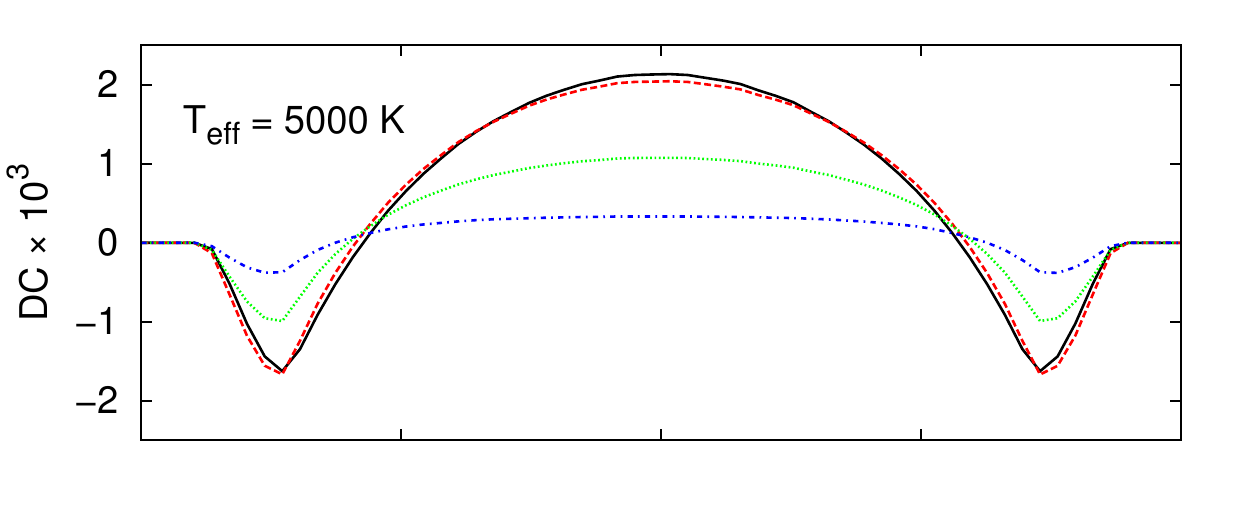}
  \\
  \includegraphics[angle=0, width=0.49\textwidth]{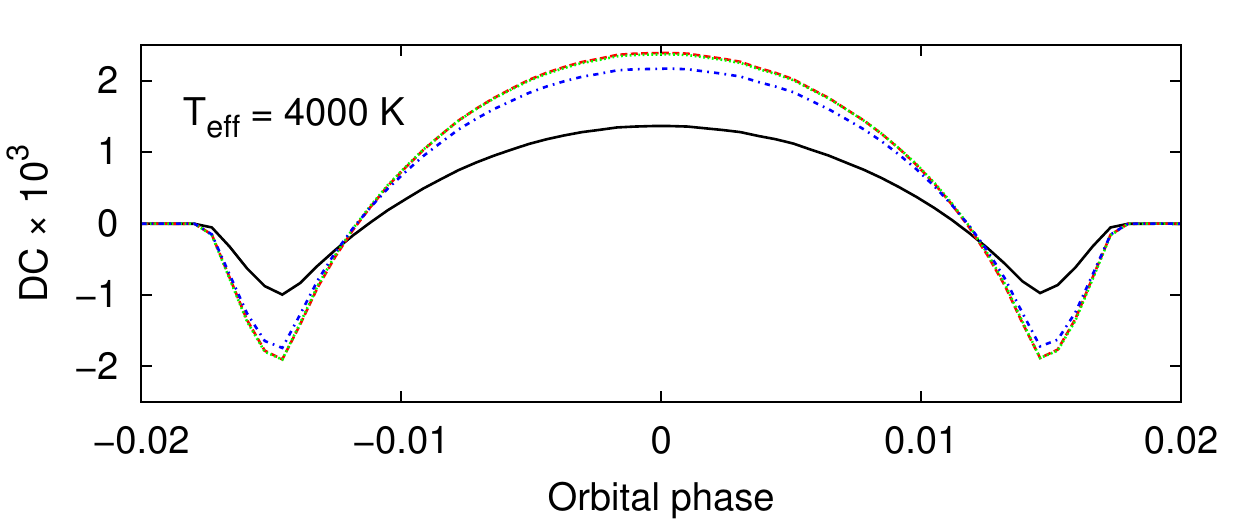}
  \includegraphics[angle=0, width=0.49\textwidth]{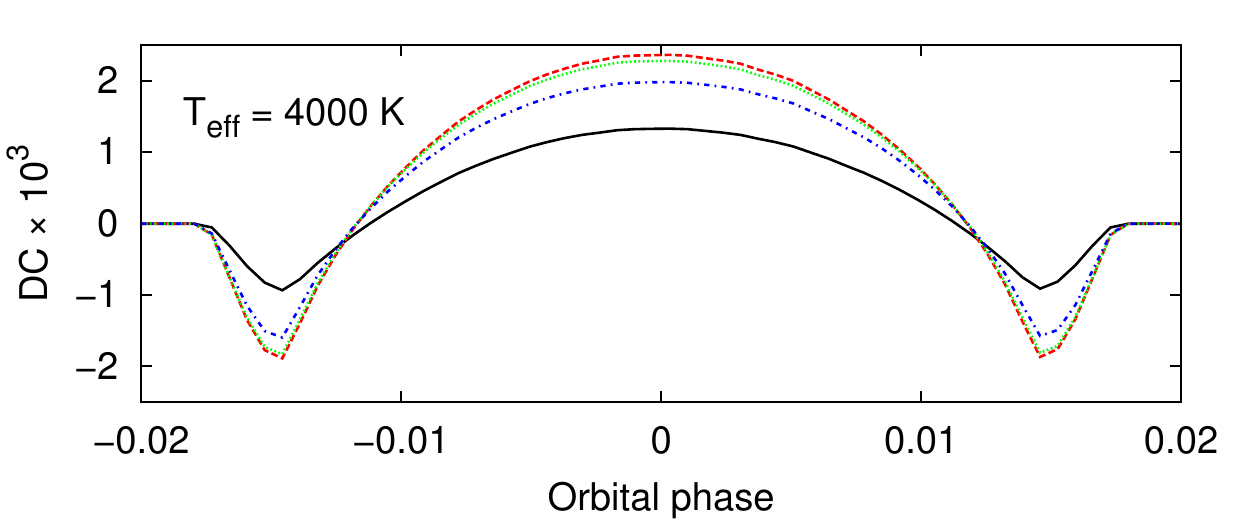}
  \caption{Simulated DCs for the \nal\ lines
  (D$_2$ in the left column and D$_1$ in the right column) for effective
  temperatures of 4000~K, 5000~K, and 6000~K. Normalized feature
  light curves where extracted in four bands with half-widths $0.1$~\AA,
  $0.375$~\AA, $0.75$~\AA, and $1.5$~\AA\ centered the respective sodium line.
  The resulting DCs are plotted using solid (black), dashed (red), dotted
  (green), and dash-dotted (blue) lines in the same order.
  \label{fig:simNaTran}}
\end{figure*}

From our simulations,
we obtained the feature light curve, $f_f(t_i)$, by integrating the spectral
signal in bands with half-widths of $0.1$~\AA, $0.375$~\AA, $0.75$~\AA, and
$1.5$~\AA, centered on either of the two sodium lines. The reference light curve
is defined as the median flux obtained across a broad ($5820-5950$~\AA)
wavelength range. The resulting DCs are shown in
Fig.~\ref{fig:simNaTran} for stars with effective temperatures of $4000$~K,
$5000$~K, and $6000$~K. The associated DCEs are listed in
Table~\ref{tab:charboNarrow}. For all stars and bands considered here, we obtain
positive DCEs. Therefore, the sodium lines show net brightening during transit.

The actual choice of a reference interval
in the simulated spectral region remains quite insignificant for the resulting
DCs. Although there is some differential limb darkening even in the
continuum,
the level of change is
much smaller than the deviations seen in the spectral lines. For instance, the DC of the two
potential reference intervals $5860-5865$~\AA\ and $5920-5940$~\AA\ on the blue
and red side of the sodium lines shows variation in the $\pm 2\times 10^{-5}$
range for a star with $T_{\mathrm{eff}} = 4000$~K, which is two orders of magnitude
smaller than that observed in the \nal\ lines.

\begin{table}
\caption{Difference curve excesses for narrow reference bands with full
widths of $0.75$~\AA, $1.5$~\AA, and $3$~\AA\ centered on the \nadd\ lines. All
DCEs are given in units of $10^{-5}$.
\label{tab:charboNarrow}}
\begin{tabular}{l r r r r r r} \hline \hline
T$_{\mathrm{eff}}$ $[$K$]$ & \multicolumn{2}{c}{0.75~\AA} & \multicolumn{2}{c}{1.5~\AA} &
\multicolumn{2}{c}{3.0~\AA} \\
 & D$_2$ & D$_1$ & D$_2$ & D$_1$ & D$_2$ & D$_1$ \\ \hline
4000 &  78.9 &  78.1 &  78.5 &  75.6 &  72.1 &  66.1 \\
4500 &  93.3 &  94.1 &  94.0 &  88.6 &  77.2 &  57.1 \\
5000 &  72.6 &  66.8 &  54.3 &  37.1 &  25.0 &  13.1 \\
5500 &  49.8 &  36.2 &  23.9 &  13.9 &   9.9 &   5.0 \\
6000 &  30.4 &  19.4 &  11.6 &   7.1 &   5.3 &   3.0 \\
6500 &  16.0 &  10.5 &   5.7 &   3.9 &   2.7 &   1.6 \\
7000 &   9.0 &   7.2 &   3.3 &   2.9 &   1.6 &   1.2 \\ \hline
\end{tabular}
\end{table}

The DCs involving the sodium line show variability on a level of up to $10^{-3}$.
The behavior is typical for transit light curves obtained from a
limb-brightened feature (cf. Sect.~\ref{sec:diffandrat}). The details obviously
depend on the choice of the extraction bands and CLV of
the stellar lines. Nonetheless, the amplitude of the curves shown in
Fig.~\ref{fig:simNaTran} can clearly be sufficient to be relevant for the study
of exoplanetary atmospheres, whose signals are expected to be on the order of
$10^{-3}-10^{-4}$ \citep[e.g.,][]{Seager2000}.

\subsubsection{Treatment of the CLV-induced in-transit excess}
Through the partial occultation of the stellar disk during transit,
the stellar CLV is imprinted on the DC. In the cases considered here, it causes
apparent net emission in the feature bands covering the \nadd\ lines. This has
to be taken into account in the measurement of in-transit excess absorption
due to the planetary atmosphere as has been shown, e.g., by
\citet{Charbonneau2002} and \citet{Sing2008LD}. In particular, the corresponding
value should be added to the measured transit depth, if we have not previously taken the CLV  into account.

As the DC is time dependent, the result of the averaging required to obtain the
DCE (see Sect.~\ref{sec:naming}) depends on the sampling of the transit. An
accurate correction of the CLV-induced effect is only possible, if all boundary
conditions like the temporal sampling, instrumental resolution, and sensitivity
are taken into account in the modeling.

\section{Observation of the center-to-limb variation in \hde}
\label{sec:Observations}

In 2012, we obtained high-resolution transit spectroscopy of \hde\
with UVES. The properties of \hde\ are summarized in Table~\ref{tab:Properties}.
We generated synthetic spectra of the star based on a Kurucz model
atmosphere with an effective temperature of $5000$~K, a surface gravity of
$\log(g)=4.5$, and solar abundances.
Again, we used our code to simulate a spectral time series based on the
orbital configuration of \hde\ system from which we derive model
light curves to be compared with the observations.
 
\begin{table}
  \centering
  \caption{Properties of \hde.
  \label{tab:Properties}}
  \begin{tabular}[h]{l l l}
  \hline\hline
  Property & Value & Source$^a$ \\ \hline
  \multicolumn{3}{c}{The host star} \\
  $T_{\mathrm{eff}}$ [K] & $5040\pm 50$ & T \\
  $\log$(g[cm\,s$^{-2}$]) & $4.587^{+0.014}_{-0.015}$ & T \\
  $[$Fe/H$]$ & $-0.04\pm 0.08$ & B/T \\
  $v\sin(i)$ [km\,s$^{-1}$] & $3.5\pm 1$ & B \\
  M$_{\mathrm{s}}$ [M$_{\odot}$] & $0.806\pm0.048$ & T \\
  R$_{\mathrm{s}}$ [R$_{\odot}$] & $0.756\pm0.018$ & T \\
  \multicolumn{3}{c}{The planet and its orbit} \\
  SMA$^b$ [R$_{\mathrm{s}}$] & $8.81\pm0.06$ & T \\
  R$_{p}/$R$_{s}$ & $0.15463\pm0.00022$ & T \\
  $i_{\mathrm{orbit}}$ [deg] & $85.58\pm0.06$ & T \\
  $T_0$ [BJD$_{\mathrm{UTC}}$] & 2454279.436714 & A \\
  period [d] & 2.21857567 & A \\
  \hline
  \end{tabular}
  \tablefoot{$^a$\; B:\citet{Bouchy2005}, T:\citet{Torres2008}, A:\citet{Agol2010};
  $^b$\; Orbital semi-major axis}
\end{table}

\subsection{Observations and data reduction}
\label{sec:obsandred}
On July 1, 2012, we observed \hde\ with UVES mounted at the VLT-UT2 (Kueyen).
Between 04:06 and 08:40~UT, we obtained $244$ spectra,
whereof the first $29$ were exposed for $30$~s and the remaining for $45$~s.
We applied the dichroic beam splitter (dic2) in the
$437+760$~nm setup. The slit width was $1''$ for the blue and $0.7''$ for the
red arm. To optimize the temporal cadence, we used the ``ultra fast readout''
option.

To reduce the echelle spectra, we used the UVES-pipeline in version
$5.2.0$. The pipeline performs bias subtraction, hot-pixel rejection, order
definition, order extraction, flat fielding, and the wavelength calibration
using ThAr frames.
Finally, the echelle orders are merged into a single 1d-spectrum and flux
calibrated. In this work, we use the spectra from the blue and
lower red chip, which were obtained using optimal extraction.

\begin{figure}[h]
  \includegraphics[angle=0, width=0.49\textwidth]{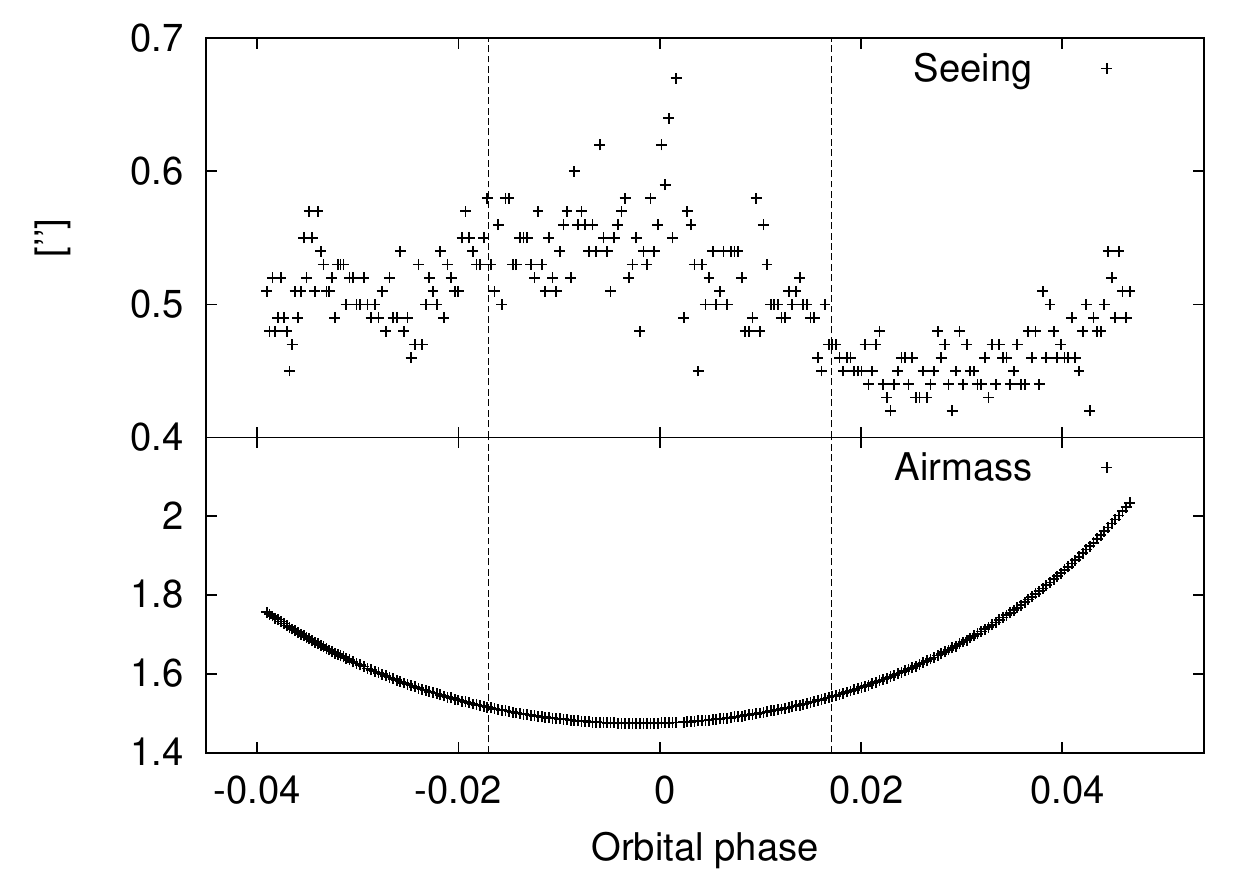}
  \caption{Temporal evolution of seeing (top) and air mass (bottom). The vertical
  dashed lines indicate the duration of the transit. The temporal offset, $T_0$,
  refers to $\mbox{MJD}_{HJD,UTC} = 56109.26164$.
  \label{fig:SeeingAirmass}}
\end{figure}

We obtained reference spectra for flux calibration during our
observing night, but found that the spectra calibrated using the master response
files\footnote{see
\url{http://www.eso.org/observing/dfo/quality/UVES/qc/SysEffic_qc1.html}.
We used \texttt{UV\_MRSP\_041130\_BLUE437} for the
blue, \texttt{UV\_GMRE\_070511A\_master\_response\_REDL760} for the lower red,
and \texttt{UV\_GMRE\_090701A\_master\_response\_REDU760} for the upper red
chip.}
show a lower degree of large-scale structure not attributable to the
stellar spectrum, while the small-scale noise properties
of the resulting spectra are identical. The flux obtained with the
master response files was generally lower by about $15-20$\% on the lower red
chip, which gives an idea of the involved uncertainties.
In our analysis, we prefer to use the spectra calibrated using the master
response files.

The blue and red chip are read out separately and, in principle, asynchronously.
Analyzing the start times as
determined by the header keyword \texttt{MJD-OBS}, we find that, on average, the
observation obtained with the blue chip starts $0.83\pm 1.1$~s later than on the red chip.
In the most extreme case, the observation obtained with the blue chip is delayed by $9$~s
compared to the red chip. As the difference is, however, larger than $3$~s
in no more than six cases, the observations can be treated as simultaneous
where not explicitly stated otherwise.

The data cover  the primary transit plus some pre- and post-transit time. In
Fig.~\ref{fig:SeeingAirmass}, we show the temporal evolution of the seeing
(header keyword \texttt{HIERARCH ESO TEL IA FWHM}) and the air mass during our
observations. The duration of the transit is explicitly
indicated. The seeing is between $0.5$'' and $0.6$'' during the first half of
the observations and further improves in the second half. 

\subsection{Accuracy of the flux-calibration}
\label{sec:fluxCal}
The flux calibration applied by the UVES pipeline takes  the
exposure time, gain, binning, atmospheric extinction, and
air mass\footnote{``UVES Pipeline User Manual'' Issue 22.3, Sect. 11.1.21} into account.
To verify that reasonable fluxes are produced, we integrated the flux obtained
with the individual chips and compare the numbers thus obtained with a model
spectrum and a black body. The model spectrum is based on a Kurucz
model atmosphere with an effective temperature of $5000$~K, a $\log(g)$ value
of $4.5$, and solar metallicity \citep{Castelli2004, Kurucz1970}. The spectrum
was synthesized using the \texttt{spectrum} program \citep{Gray1994}. To
calculate the blackbody spectrum, we applied a temperature of $5040$~K (see
Table~\ref{tab:Properties}).
In Table~\ref{tab:fluxCal}, we list the minimum and maximum observed flux on
each chip along with the flux predicted by the Kurucz atmosphere and the
black body.

\begin{table}[h]
  \caption{Measured and predicted fluxes at Earth.
  \label{tab:fluxCal}}
  \begin{tabular}{l l l l l} \hline\hline
  Chip & Wvl. range & Obs. flux & K$^a$ & BB$^a$ \\
       & [$\AA$]   & \multicolumn{3}{c}{[$10^{-9}$ erg\,cm$^{-2}$\,s$^{-1}$]} \\
  \hline
  blue & $3732.1 - 4999.7$ & $2.3 - 2.8$ & $2.9$ & $3.3$ \\
  redl & $5655.1 - 7595.1$ & $4.8 - 6.0$ & $5.6$ & $5.9$ \\
  redu & $7564.3 - 9463.9$ & $3.5 - 4.2$ & $4.0$ & $4.4$ \\ \hline 
  \end{tabular}
  \tablefoot{$^a$Kurucz atmosphere (K), black body (BB).}
\end{table}

\begin{figure}[h]
  \includegraphics[angle=0, width=0.49\textwidth]{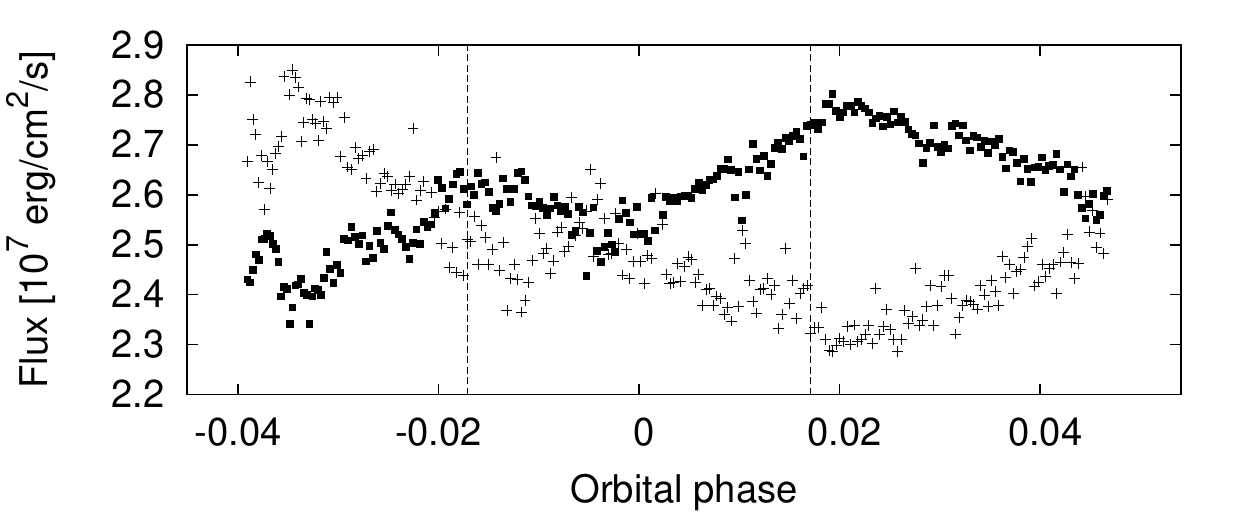}
  \caption{Temporal evolution of the measured spectral flux on the blue chip
  (squares) and (scaled) signal in the acquisition images from the blue camera
  (crosses). Dashed vertical lines indicate the transit duration.
  \label{fig:sigVsTime}}
\end{figure}

In Fig.~\ref{fig:sigVsTime}, we show the
measured spectral flux on the blue chip obtained by integrating the
flux-calibrated spectrum and the signal measured in the blue-channel acquisition
images as a function of time. The latter clearly represents light having
missed the slit. Acquisition and spectral signal are strongly anticorrelated.
The evolution of the spectral and acquisition signal after the first half
of the observation is clearly related to the behavior of the seeing
(Fig.~\ref{fig:SeeingAirmass}) with better seeing resulting in more stellar
light entering the slit.

Although the spectral signal shows an apparent drop, which is remarkably
well aligned with the transit, the depth of this hypothetical transit is
about 5\% and, therefore, it is too deep to be (solely) caused by the planet, 
which shows
a transit depth of about $3$\% in the B filter \citep{Bouchy2005}. While the
spectral and acquisition signals can be combined to obtain a curve showing a
transit of proper depth, the uncertainty in the scaling prevents a meaningful
quantitative analysis.

The observed flux and the prediction based on the Kurucz atmosphere agree on the
20\% level, which seems reasonable considering the systematic
uncertainty in the flux calibration (see Sect.~\ref{sec:obsandred}).
The residual variation is largely attributable to
slit losses with some residual influence of 
air mass and atmospheric conditions that are likely also present.

\subsection{Wavelength stability}
\label{sec:wavestability}

In the red part of the spectrum ($\lambda \gtrsim 6000$~\AA), telluric lines
yield a radial velocity calibration with an accuracy of about $20$~\ms\
\citep[e.g.,][]{Balthasar1982,Caccin1985, Gray2006}.
We used the telluric lines to study the stability of the wavelength calibration 
in
our data. First, we obtained a nominal telluric transmission spectrum using the
Line-By-Line Radiative Transfer Model (LBLRTM) code \citep{Clough2005}. Second, we
selected a total of six wavelength intervals between 6800 and 9200~\AA\footnote{
We used the intervals $6864-6968$~\AA, $7160-7221$~\AA,
$7221-7351$~\AA, $8128-8180$~\AA, $8949-9037$~\AA, and $9055-9138$~\AA.}, and
third, we determined the (apparent) radial velocity shift of the telluric lines using
a cross-correlation.

\begin{figure}[h]
  \includegraphics[angle=0, width=0.49\textwidth]{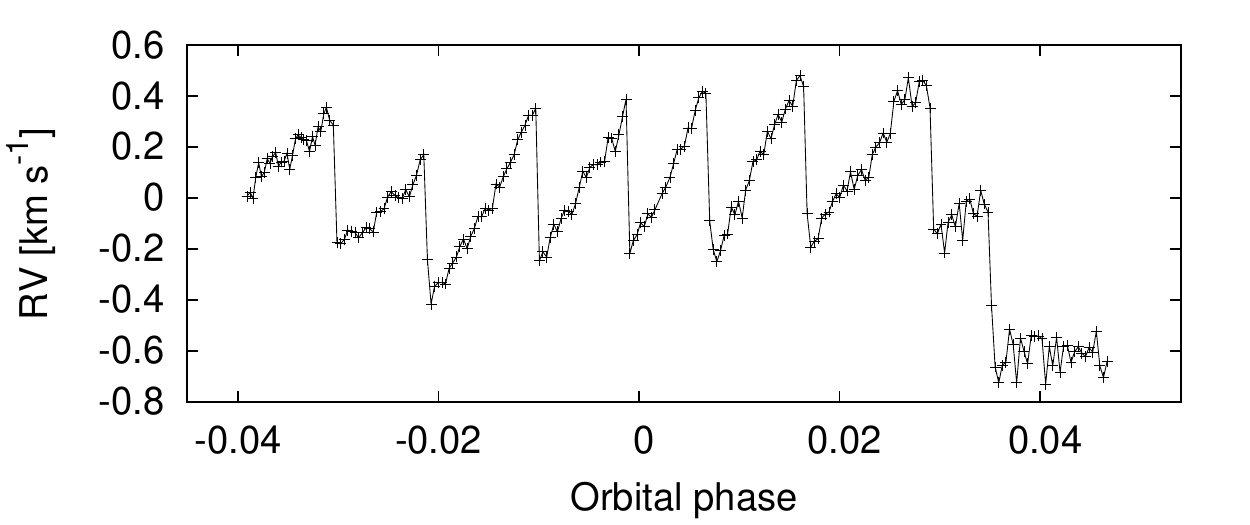}
  \caption{Radial velocity shift of telluric lines between $6864$ and
  $6968$~\AA.
  \label{fig:tellurRV}}
\end{figure}

Figure~\ref{fig:tellurRV} shows the resulting shift of the telluric oxygen lines
between $6864$ and $6968$~\AA. The curve shows a sawtooth-like
pattern with an overall amplitude of roughly 1~\kms, which we found for telluric
lines across the entire red spectral arm. We attribute the
variation to instrumental effects, mainly changes in the positioning of the
stellar disk within the slit, which also manifest in a shift of the location of
the echelle orders on the detector and changes in the asymmetry of the
acquisition images. The slit width of 0.7'' in the red arm was,
indeed, larger than the seeing disk; the same holds for the blue arm.

To correct for the instrumental shifts, taking  its wavelength
dependence into account, we used the telluric corrections determined in the six individual
bands and interpolated to correct at intermediate wavelengths. 

\subsection{The \rme}
The \rme\ is an apparent radial velocity shift caused by a change in the
spectral line profile during the planetary transit
\citep[e.g.,][]{Rossiter1924, McLaughlin1924}.
The Rossiter-McLaughlin curve provides important
information about the planetary system, most notably,
the relative orientation of the (sky-projected) stellar spin axis and the 
planetary orbit normal
\citep[e.g.,][]{Ohta2005}.

\begin{figure}[h]
  \includegraphics[angle=0, width=0.49\textwidth]{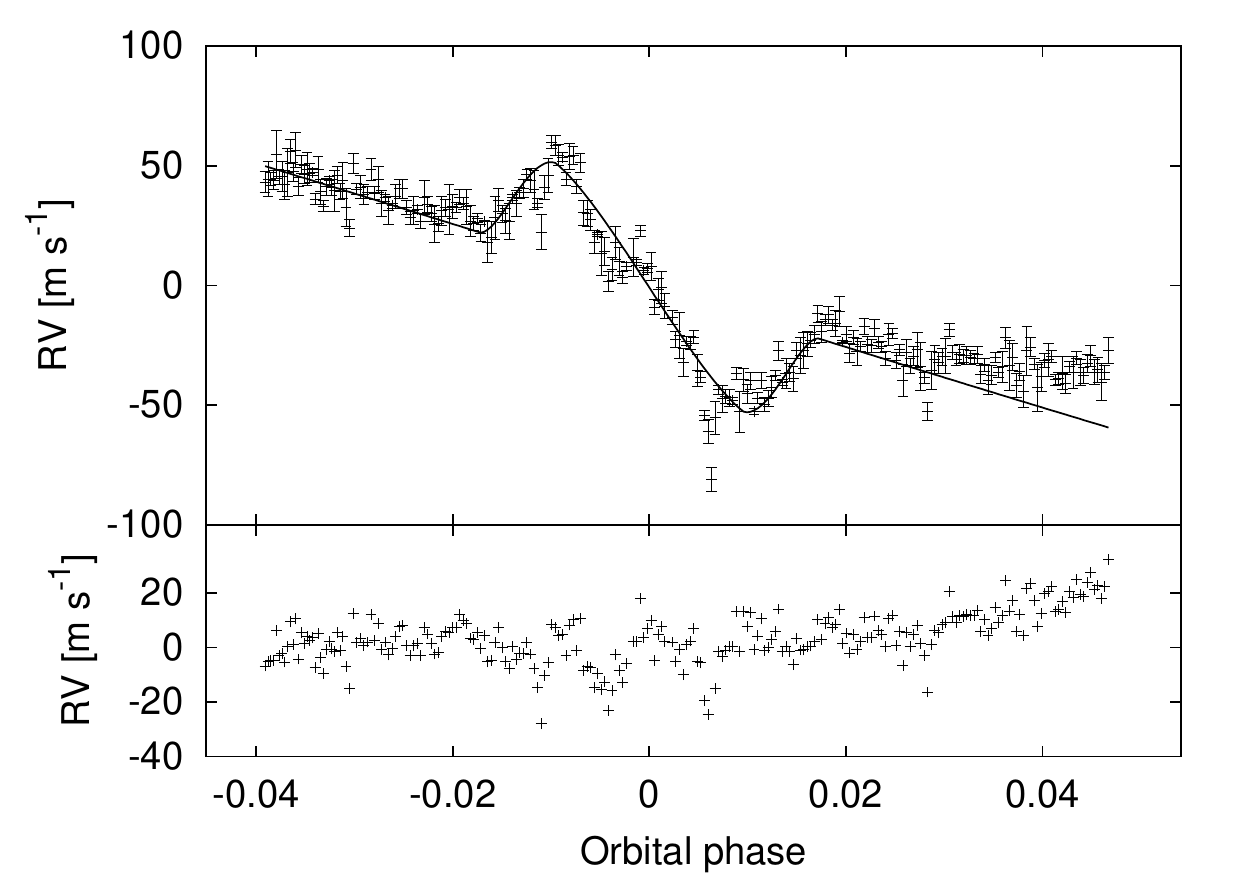}
  \caption{Rossiter-McLaughlin radial velocity curve along with nominal model
  with parameters from \citet{Triaud2009} (top) and residuals (bottom).
  \label{fig:RmcL}}
\end{figure}

In \hde, the amplitude of the \rme\ is about $40$~\ms, which is a factor of 25
below the instrumental velocity shifts seen in the telluric lines. To measure
the \rme, we determined the stellar radial velocity by cross-correlation with a
synthetic spectral model in the $6620-6700$~\AA, $6700-6840$~\AA,
$7780-7820$~\AA, and $8700-8750$~\AA\ bands.
The stellar radial velocity was then corrected by subtracting the telluric
correction determined in Sect.~\ref{sec:wavestability} and adding the
barycentric correction.
The spectral bands used in the cross-correlation were selected because
they are close to the telluric bands used to correct the instrumental radial
velocity shift, but are themselves only weakly affected by telluric absorption.
The resulting curve is
shown in Fig.~\ref{fig:RmcL} along with a nominal model calculated using the 
parameters
from \citet{Triaud2009, Ohta2005}. The errors were estimated from the scatter of
the radial velocity measurements obtained in the individual spectral bands. Note
that we did not carry out a fit.

The \rme\ is clearly seen. Toward the end of the observation run, there seems to
be a systematic offset between model and measurement, which may due to a
shortcoming of the instrumental correction. Otherwise, the \rme\ is compatible
with the curve reported by \citet{Triaud2009}. During the transit, we obtain
correlated residuals not unlike those seen in previous works
\citep[see][]{Winn2006, Triaud2009}. The residuals seen by \citet{Winn2006} and
\citet{Triaud2009} show a remarkable point symmetry with respect to mid-transit
time, which is also indicated by our residuals.

\subsection{Line profiles and instrumental resolution}
\label{sec:instrResol}
\begin{figure}[h]
  \includegraphics[angle=0, width=0.49\textwidth]{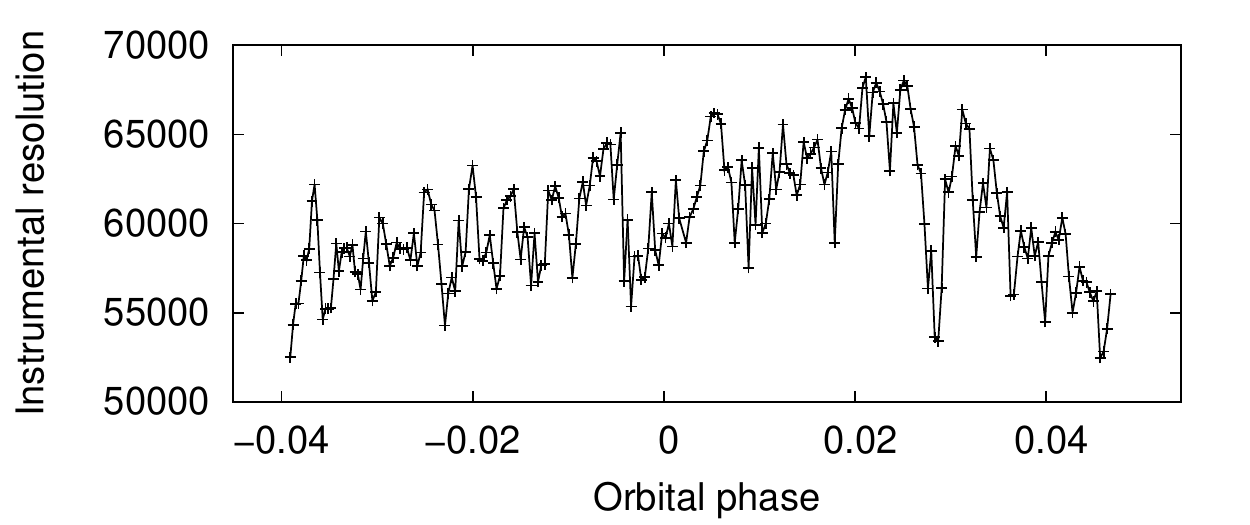}
  \caption{Equivalent-width weighted instrumental resolution obtained from $19$
  telluric lines between $5885$ and $7265$~\AA.
  \label{fig:meanFWHM}}
\end{figure}

Changes in the line profile potentially interfere with all analyses based on
individual line characteristics, in particular, the line depth or flux
observed around the center of the line.
To study the temporal stability of the line profile, we selected
a total of 19 sufficiently isolated, telluric spectral lines between $5885$~\AA\
and $7265$~\AA. All the lines were modeled using a Gaussian profile. Assuming
that the lines are intrinsically narrow, the instrumental resolution is obtained
by the inverse of the FWHM times the wavelength.
Figure~\ref{fig:meanFWHM} show the equivalent-width weighted mean instrumental
resolution. The temporal evolution of the resolution is anticorrelated with the
seeing (see Fig.~\ref{fig:SeeingAirmass}).

According to our analysis, the mean instrumental resolution is about $60\,000$,
but it clearly varies during the observing run.
The behavior can likely be attributed to a change in slit illumination, with the
seeing disk moving in the slit.

\section{Center-to-limb variation observed in the  \cahk\
lines}
\label{sec:CLV_CaHK}
The \cahk\ lines at 3933.66~\AA\ and 3968.47~\AA\ are the widest
and strongest spectral lines in the wavelength range covered by our UVES observations.
These lines are expected to show a distinct pattern of CLV across their
profiles, and their width makes them favorable targets for studying the CLV.
Prior to any further analysis, we cross-correlated the observed spectra
with a model spectrum to correct for the instrumental radial-velocity shift
(see Sect.~\ref{sec:wavestability}). In the following,
we adopt the reference continuum
bands $C_{\mathrm{H}}$ ($3891.67-3911.67$~\AA) and
$C_{\mathrm{K}}$ ($3991.067-4011.067$~\AA) also used by \citet{Melo2006}.

\subsection{The line core}
\label{sec:cahkCoreDC}
\begin{figure}
  \includegraphics[angle=0, width=0.49\textwidth]{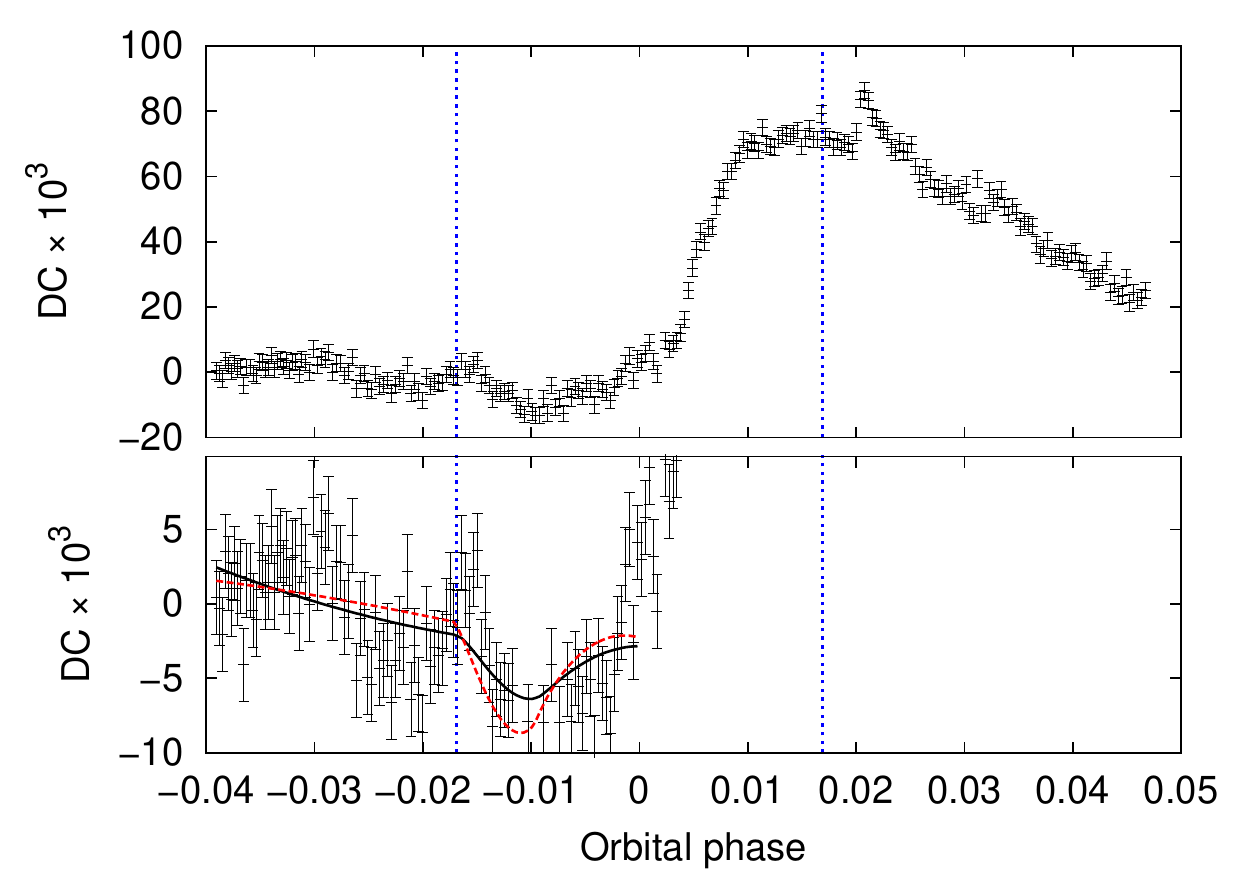}
  \caption{Top: the observed DC extracted from the \cahk\ line cores. Bottom:
  close-up of the first half of the observed DC. The solid line shows a
  photospheric model DC and the dashed line a chromospheric model DC, assuming
  homogeneous distribution of the \cahk\ core emission.
  \label{fig:cahkcore}}
\end{figure}

The cores of the \cahk\ lines are highly sensitive to chromospheric emission
\citep[e.g.,][]{Baliunas1995}. To obtain the DC,
we integrated the
spectral signal in $1$~\AA\ wide bands centered on the \cahk\
lines cores and derived two normalized feature light curves $LC_K(t_i)$ and
$LC_H(t_i)$. The integration of the signal in the two aforementioned reference
bands yields two normalized reference light curves, $LC_{C_K}(t_i)$ and
$LC_{C_H}(t_i)$. Based on these light curves, we obtained the DC as
\begin{equation}
  DC_{core}(t_i) = \frac{(LC_K(t_i) + LC_H(t_i))}{2} - \frac{(LC_{C_K}(t_i) +
  LC_{C_H}(t_i))}{2} \; ,
  \label{eq:DCfromObs}
\end{equation}
where the first term is the mean feature light curve and the second the mean
reference light curve. In Fig.~\ref{fig:cahkcore}, we show the resulting DC,
which shows a complex temporal pattern of variability. Most
prominently, a steep rise is seen at about mid-transit time, which is followed
by a decay phase, lasting for the rest of the observation. This behavior in
chromospherically sensitive lines is typical of solar and stellar flares
\citep[e.g.,][]{Fang1992, Fuhrmeister2008}. Therefore, we attribute it
to a flare. Superimposed on the
decay phase of the flare is another smaller, yet otherwise similar, event. We
interpret this as another, potentially related, flare.
Indeed, multiple temporally overlapping flares are frequently observed on
active stars \citep[e.g.,][]{Lalitha2013}. In the following analysis, we
exclude the second half of the DC from our modeling of the CLV.

To model the observed behavior in the first half of the observation, we
derived two synthetic DCs.
First, we obtained a
photospheric
% \LEt{Is this a quotation? If so, you can keep the quotation marks.}
model, representing the DC that would be obtained in the
absence of any chromospheric emission from our
synthetic spectral time series. By fitting a transit model with quadratic
limb darkening \citep{Mandel2002} to the synthetic mean feature and reference
light curves, we obtained limb-darkening coefficients of $a_f=0.27$ and
$b_f=0.33$ for the feature and $a_r=1.08$ and $b_r=-0.24$ for the reference. The
latter are compatible with the values of $a=1.06$ and $b=-0.2$ derived by
\citet{Claret2004} for stars like \hde\ in the Sloan $u'$-band. 
Second, we constructed a DC assuming a homogeneous distribution of chromospheric
emission across the stellar disk, i.e., zero limb darkening of the feature.
The reference light curve was left unchanged.
To model the observed DC, we additionally
added a second second-order polynomial to the DCs to take  the
global pattern of variability into account. The actual spatial distribution of
chromospheric emission is probably more inhomogeneous as observed on the Sun
\citep{Llama2015}, but we refrain from adopting a more complex model here.

In the bottom panel of Fig.~\ref{fig:cahkcore}, we show the first half
of the data along with the our two best-fit photospheric and chromospheric
models.
In the fitting, only the coefficients of the
polynomial were varied.
The behavior of the observed DC during ingress
is reminiscent of the expected behavior. In particular, an initial decrease
followed by an increase toward the mid-transit phase is clearly visible. The
behavior is better reproduced by the chromospheric model, assuming homogeneously
distributed
emission in the \cahk\ emission line cores. While this is plausible, there is
also variability on a similar scale as observed prior to ingress, which is,
therefore, most probably unrelated to the transit. Although the model is
plausible, it remains difficult, if not impossible, to distinguish between
intrinsic variability and CLV-induced changes in the cores of the \cahk\ lines.

\subsection{The line wings}
\label{sec:cahkWingDC}

\begin{figure}
  \includegraphics[angle=0, width=0.49\textwidth]{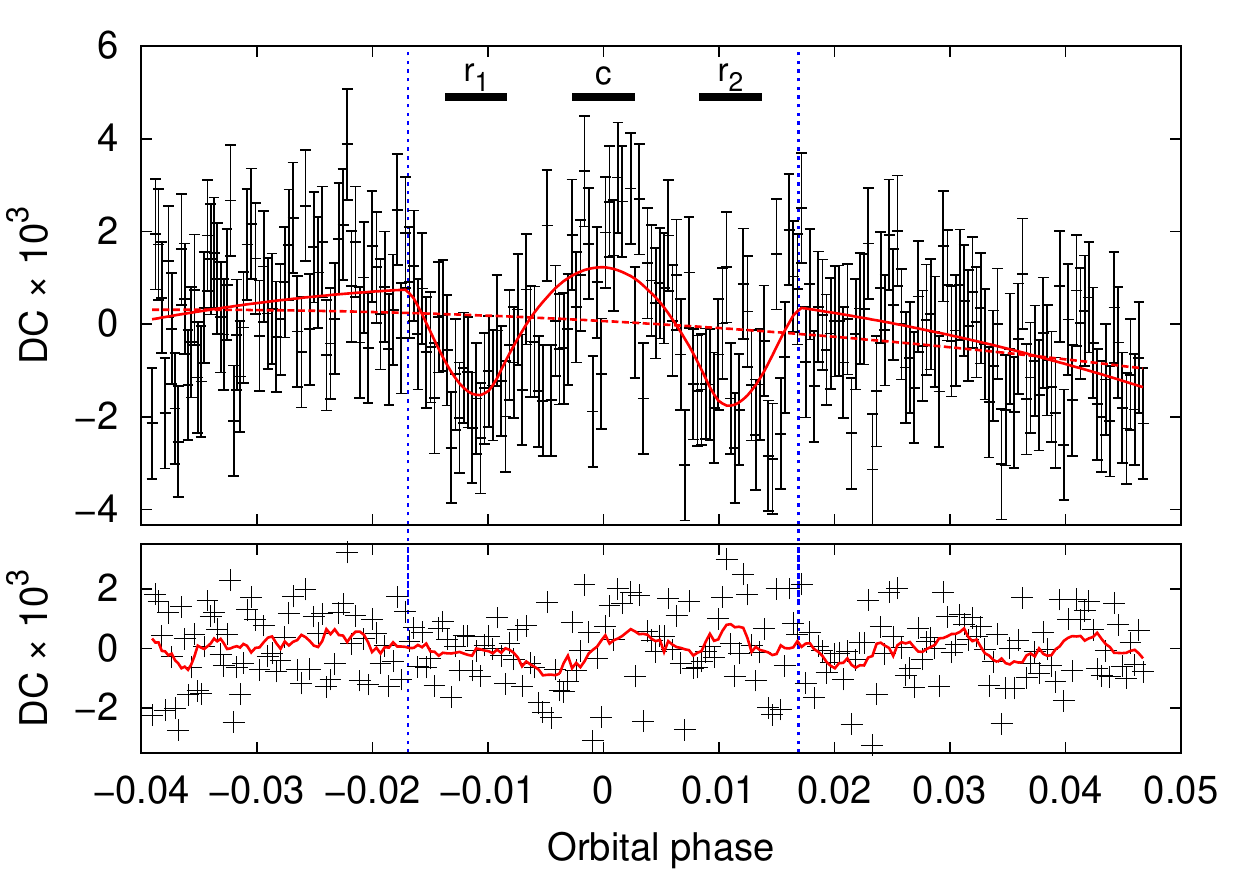}
  \caption{Combined DC of the \cahk\ lines wings along with our
  best-fit second order polynomial model (dashed) and our best-fit model including the
  synthetic DC. The bottom panel gives the residuals with respect to our DC
  model.
  The solid (red) line shows the residuals smoothed by a running mean with a
  width of eight data points. The vertical
  lines indicate the transit duration and the three top bars indicate the phase
  intervals used in the analysis of the spectra.
  \label{fig:cahkclv}}
\end{figure}

In contrast to the line cores, the wings of the \cahk\ lines are not expected to
be heavily affected by stellar activity. 
In Fig.~\ref{fig:cahkclv}, we show the DC derived from the
\cahk\ lines of \hde. Here, our feature light curve was extracted by integrating
the spectrum in a
region between $3$~\AA\ and $5$~\AA\ from either line core, while the cores
themselves were excluded.  A separation of $3$~\AA\ from the
\ion{Ca}{ii} H core also excludes the H$\epsilon$ line, which is not
unexpectedly prominently seen during the flare (see Fig.~\ref{fig:Heps}). We
obtained four normalized feature light curves corresponding to the $2$~\AA\ wide bands on the
blue and red side of the \cahk\ line cores and obtained the mean feature
light curve by averaging (cf. Eq.~\ref{eq:DCfromObs}).

The DC in the line wings shows the general pattern of variability expected from
a limb-brightened feature light curve (see Sect.~\ref{sec:diffandrat}). However, there
clearly is additional large-scale evolution in the DC, which is
reminiscent of the behavior of air mass (see Sect.~\ref{sec:obsandred}).
After having estimated the noise from the scatter
in the curve, we modeled the curve, first, using a second-order polynomial
and, second, using the polynomial plus a synthetic DC.
The resulting models are also shown in
Fig.~\ref{fig:cahkclv}.
For the polynomial
alone, we obtained a reduced $\chi^2$ value of $1.31$. After adding the model
DC, we obtained a reduced $\chi^2$ value of $1.02$. Because we have not added
a free parameter, the second model including the synthetic DC is clearly
superior.
The probability of obtaining a reduced $\chi^2$ value of $1.02$
or higher from random noise is $40$~\%, which underlines the formal plausibility
of the presented model. While there may be some evolution not accounted for by
our model in the residuals shown in Fig.~\ref{fig:cahkclv}, in particular,
during the first half of the transit, we consider them acceptable, given that no
DC parameters were actually fitted (only the underlying polynomial). Finally,
we checked that compatible results can be produced using various feature and
reference bands to exclude a chance finding based on our particular choice of
the integration bands.
 
\subsection{Difference spectra around \cahk}
\label{sec:diffSpecsCahk}

In our next step, we aim to trace back the pronounced signal in the DC to
changes directly observable in the spectra.
Therefore, we selected the spectra showing the largest differences as indicated
by the DCs. In particular, we defined the three orbital phase intervals $r_1 =
-0.011\pm \Delta p$, $c=0.0\pm \Delta p$, and $r_2=0.011\pm \Delta p$ with
$\Delta p = 0.0027$, which results in a length corresponding to $16$\% of the
transit duration. The interval $c$ covers the transit center and $r_{1,2}$
cover the DC minimum before and after mid-transit time; they
are indicated in Fig.~\ref{fig:cahkclv}.

We then coadded all spectra matching the phase requirements, taking  their relative radial-velocity shift into
account, and then normalized the result by the
average of the two mean flux values obtained in the $C_{\mathrm{H}}$ and
$C_{\mathrm{K}}$ reference intervals. We then subtracted the coadded
mid-transit spectrum from either of the reference spectra. Finally, we 
chose similarly phased spectra from our time series of synthetic spectra and
carried out the same analysis there to obtain a model.

\begin{figure}
  \includegraphics[width=0.49\textwidth]{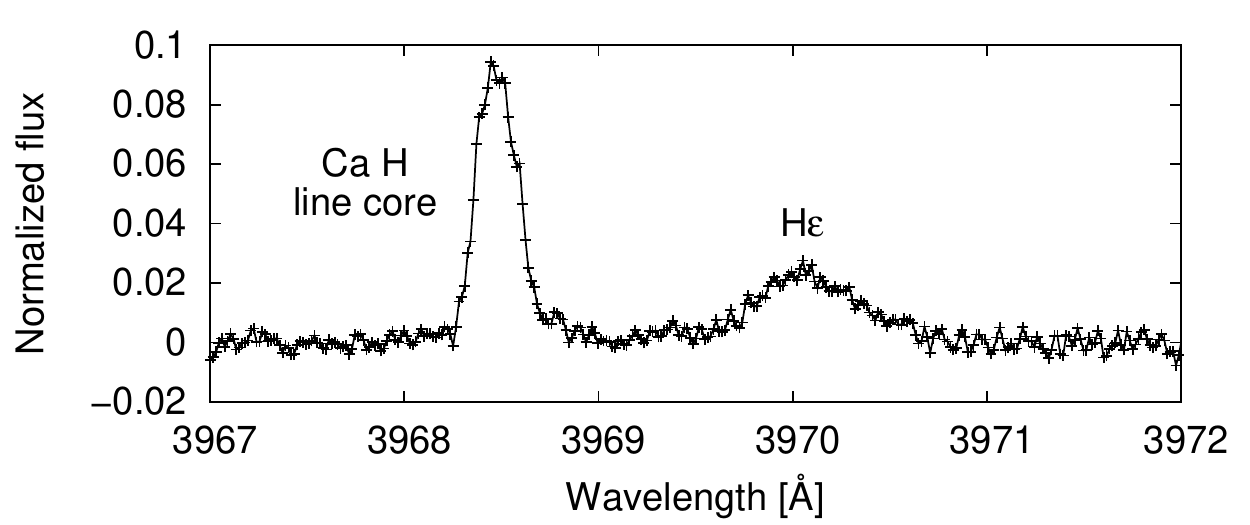}
  \caption{Close-up of the difference between the coadded spectra in the
  periods $r_2$ and $c$. The H$\epsilon$ line is clearly seen.
  \label{fig:Heps}}
\end{figure}

Figure~\ref{fig:Heps} shows a $5$~\AA\ wide excerpt of the difference
spectrum obtained by subtracting the mean spectrum in interval $c$ from the $r_2$
spectrum. As a result of the flare, a clear excess is seen in both the core of
the \ion{Ca}{ii}~H and the H$\epsilon$ line at $3970.08$~\AA.

\begin{figure}[h]
  \includegraphics[angle=0, width=0.49\textwidth]{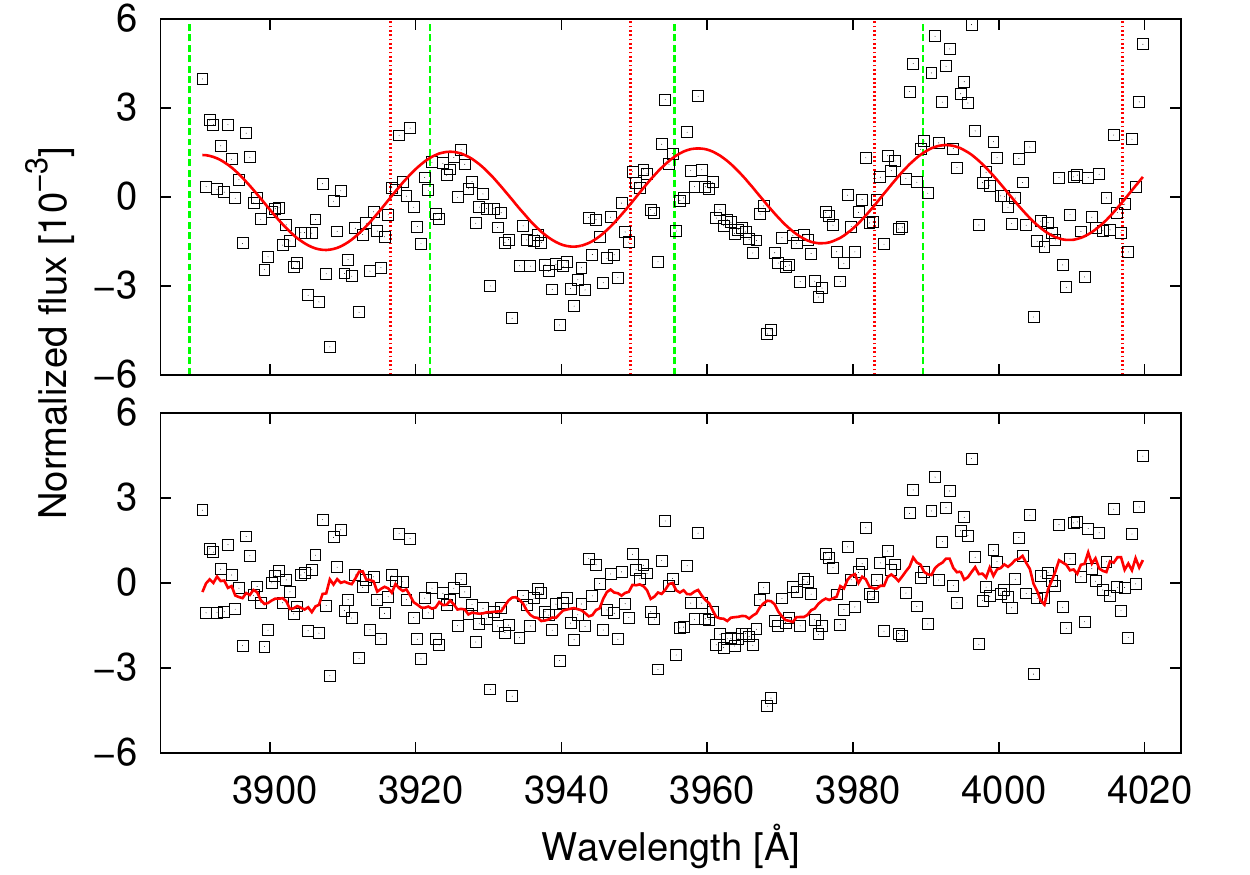}
  \caption{Top:  difference of the normalized spectra obtained in time
  intervals $r_1$ and $c$ (squares). The solid line shows the sum of our
  best-fit sinusoidal plus linear model components. 
  Vertical dashed (green) lines indicate the start and
  vertical dotted (red) lines indicate the end of an echelle order.
  Bottom: the residuals obtained after subtracting the model shown in the top
  panel from the data. The solid (red) line shows a smoothed
  synthetic difference spectrum.
  \label{fig:diffcahk_left}}
\end{figure}

On larger scales, the difference spectrum shows structure that is not
attributable to the star.
In the top panel of Fig.~\ref{fig:diffcahk_left}, we show the difference
spectrum resulting from the comparison of the coadded precenter reference
spectrum and  central spectrum.
The difference spectra were rebinned on a $0.5$~\AA\ grid to improve
visibility. Clearly, the difference spectrum shows a periodic,
sine-like signal,
whose period of $\approx 34$~\AA\ is compatible with the spacing of the UVES
echelle orders in this spectral region for our instrument setup.
Therefore, we argue that this signal is probably
related to the blaze correction and order merging of the spectra.

\begin{table}
\centering
\caption{Best-fit parameters for fit of difference spectrum
\label{tab:parsDiffSpec}}
\begin{tabular}{l l l} \hline\hline
Parameter & \multicolumn{2}{c}{Best-fit values} \\ 
 & $r_1$ - $c$ & $r_2$ - $c$ \\ \hline
$A$ & $1.6\times 10^{-3}$ & $2.5\times 10^{-4}$ \\
$\phi$ & 1.82 & 1.82 (fixed) \\
$x_0$ & $1.5\times 10^{-4}$ & $-2.3\times 10^{-4}$ \\
$g$ [\AA$^{-1}$] & $3.3\times 10^{-6}$ & $-9.3\times 10^{-6}$ \\ \hline
\end{tabular}
\end{table}

Close inspection of the synthetic and observed difference spectrum shows that
many details are not well reproduced. It must be noted that UVES resolved
hundreds of spectral lines in the region, which all show a distinct CLV.
To test whether the overall behavior of the observed difference spectrum matches
the simulation nonetheless, we smoothed the synthetic
difference spectrum using a running mean with a window size of $2.5$~\AA\ and
compared the smoothed spectrum to the observation. In particular, we fitted the
observed difference spectrum using a model, which is the sum of a sinusoidal
component, a linear component, and the synthetic difference spectrum, i.e.,
\begin{equation}
  r(\lambda) = A \sin(2\pi\nu + \phi) + x_0 + g\times (\lambda-4000~\AA) +
  \mbox{SYD}
.\end{equation}
Here, SYD represents the smoothed synthetic difference spectrum. 
The linear and sinusoidal model components represent changes unrelated to the
occultation of the stellar disk. After fixing the period, $\nu^{-1}$, of the
sine to $34$~\AA\ (the order spacing), we were left with four fit
parameters: the constant, $x_0$, the gradient, $g$, the
amplitude, $A$, and the phase, $\phi$, of the sine. Our best-fit parameters are
given in Table~\ref{tab:parsDiffSpec}. During the fit, we excluded the $\pm
2$~\AA\ wide region around the \cahk\ line cores and the same region centered on
the H$\epsilon$ line, which is clearly seen during the flare
(see Fig.~\ref{fig:Heps}).

The solid curve in the top panel of Fig.~\ref{fig:diffcahk_left} shows the
sum of the linear and sinusoidal components from our best-fit model along with
the observed difference spectrum. In the bottom panel, we plot the residuals
with respect to the curve shown in the top panel along with the smoothed
synthetic spectrum, which is the last model component. While the scatter remains
large, the overall behavior follows the expectation.

\begin{figure}[h]
  \includegraphics[angle=0, width=0.49\textwidth]{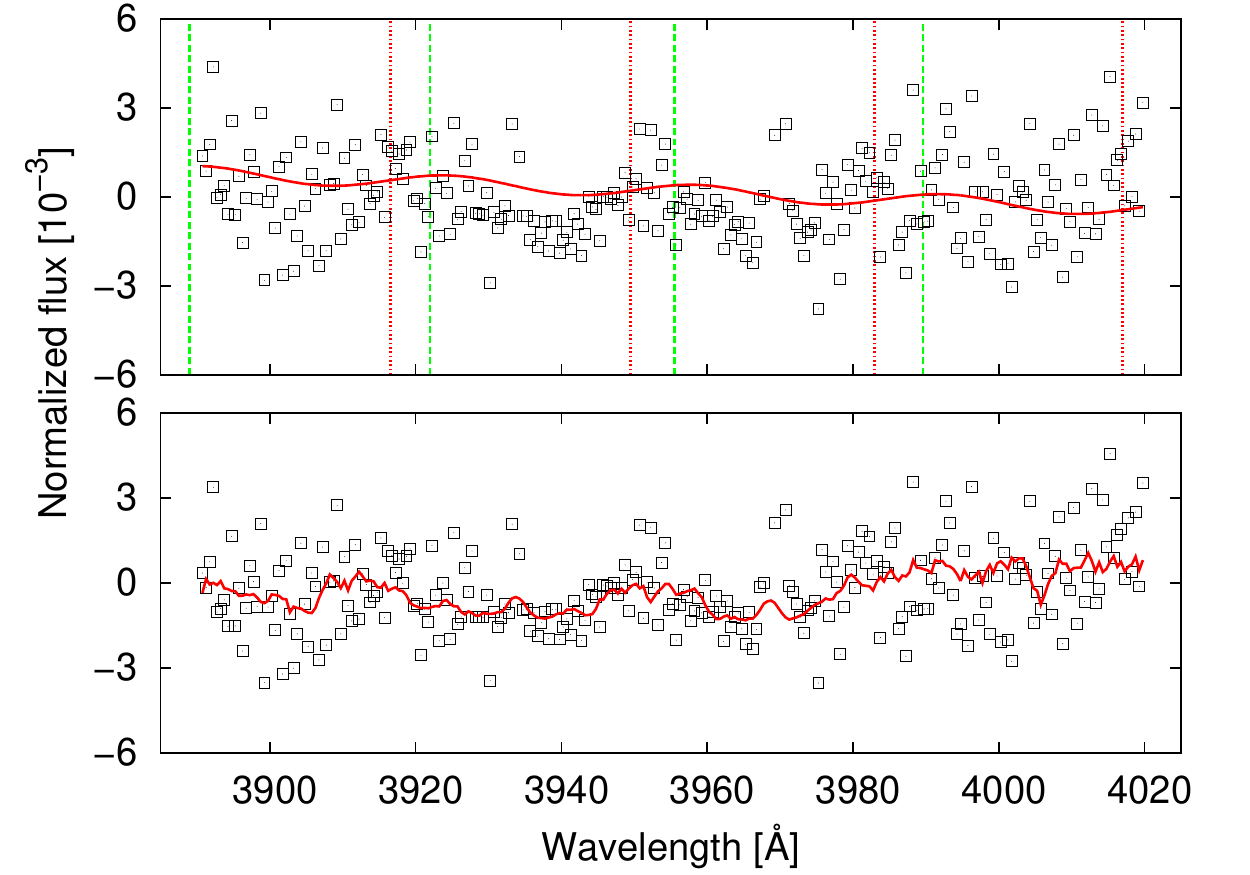}
  \caption{As in Fig.~\ref{fig:diffcahk_left} for the time intervals
  $r_2$ and $c$.
  \label{fig:diffcahk_right}}
\end{figure}

In Fig.~\ref{fig:diffcahk_right}, we show the result of the same experiment
using the post-center reference spectrum. Again, we fitted the same model to the
residuals. As the sinusoidal component is less pronounced here and we argue that
it is related to the order merging, we fixed the phase.
In this case, we obtain a stronger linear component (see
Table~\ref{tab:parsDiffSpec}), which we attribute to a change in the
atmospheric transparency not completely removed by the flux calibration.

\subsection{Influence of the systematic effects in the difference curve}

The analysis of the difference spectra in Sect.~\ref{sec:diffSpecsCahk}
clearly demonstrated the presence of systematic effects in the spectra.
To quantify the impact of these effects on the DC derived in
Sect.~\ref{sec:cahkWingDC}, we simulated spectra, including both a sine wave
with variable amplitude and a linear term with variable gradient. The
phase of the sine was fixed to the value obtained from the fit to the
residuals because this is not expected to change for an effect
related to the echelle orders. Based on these simulations, we quantify the
impact on the DC
\begin{equation}
\frac{\Delta DC}{\Delta A_{sine}} = -0.04 \;\;\; \mbox{and} \;\;\; \frac{\Delta
DC}{\Delta g} = -54~\AA \; .
\end{equation}
Given the measured amplitude of $1.5\times
10^{-3}$ of the sine wave, we estimate a resulting change on the order of
$10^{-4}$ in the DC. The expected change in the DC resulting from the gradient
is $54~\AA \times 10^{-5}~\AA^{-1} = 0.54\times 10^{-3}$, where
$10^{-5}~\AA^{-1}$ is on the order of the observed value. The change in the
gradient should, however, be slow because it is probably related to a change
in air mass. We speculate that the global trend seen in the DC, shown in
Fig.~\ref{fig:cahkclv}, is related to this effect.

Although there are instrumental effects, we argue that their magnitude
remains insufficient to explain the observed DC (see Fig.~\ref{fig:cahkclv}).
Furthermore, the temporal association with the transit seen in the DC makes an
instrumental origin unlikely. In principle, nonlinearity of the CCD could
cause an effect
temporally associated with the transit because the star appears necessarily
darker when the planet traverses the stellar disk.
% \LEt{If this isn't correct, please specify what "it" refers to here.}
However, the signal seen in the DC is not proportional to the
overall light variation expected during the transit and the signal observed on
the CCD is on the order of $1000$~ADUs in the region of the \cahk\ lines, which
is far below the saturation limit of $65\,000$~ADUs. Therefore, we argue that
nonlinearity of the CCD is unlikely to be the origin of the observed change in
the DC.

\section{Center-to-limb variation observed in the \nadd\ lines}
\label{sec:CLV_Nadd}

\begin{figure}
   \includegraphics[width=0.49\textwidth]{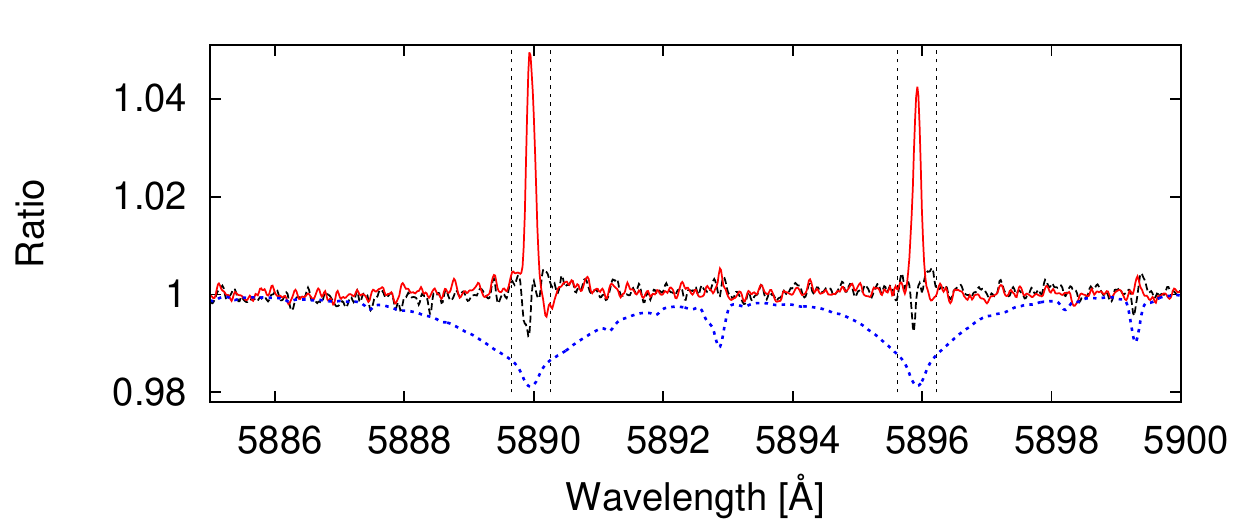}
   \caption{The solid (red) line shows the ratio of the average
   peak-flare spectrum (orbital phases $0.01$ to $0.02$) and the first out-of-flare
   reference spectrum (phase $-0.02$ to $-0.01$). For comparison, the dashed
   (black) lines shows the ratio of the averaged spectra obtained during the
   second and first out-of-flare reference phases, which shows no strong
   change. The dotted (blue) line shows the scaled stellar spectrum for
   reference.
   \label{fig:nadiffspec}}
\end{figure}

Our analysis of the \cahk\ lines shows flaring activity affecting the
line cores from about mid-transit time. To study the effect on the \nadd\
lines, we created
mean spectra in three time intervals covering orbital phases from $-0.02$ to
$-0.01$, $-0.01$ to $0.0$, and $0.01$ to $0.02$. While the first two intervals
do not cover the flare, the third interval covers its peak phase as
observed in the \cahk\ lines (cf. Fig.~\ref{fig:cahkcore}). In
Fig.~\ref{fig:nadiffspec}, we plot the ratios of the averaged spectra in the
three intervals. While the \nadd\ line cores show a clear fill-in during the
flare, the ratio of the out-of-flare spectra shows no strong change in the core
region. The core is probably also affected by the planetary
atmosphere \citep[e.g.,][]{Wyttenbach2015}, which we do not consider
here.
To minimize the impact of the flare on our CLV analysis, we excluded the line
core. In particular, we disregarded the central $\pm 0.3$~\AA\ region of the
lines, where the line profile is dominated by the central Gaussian component.
The region is also indicated in Fig.~\ref{fig:nadiffspec}.

\begin{figure}[h]
  \includegraphics[width=0.49\textwidth]{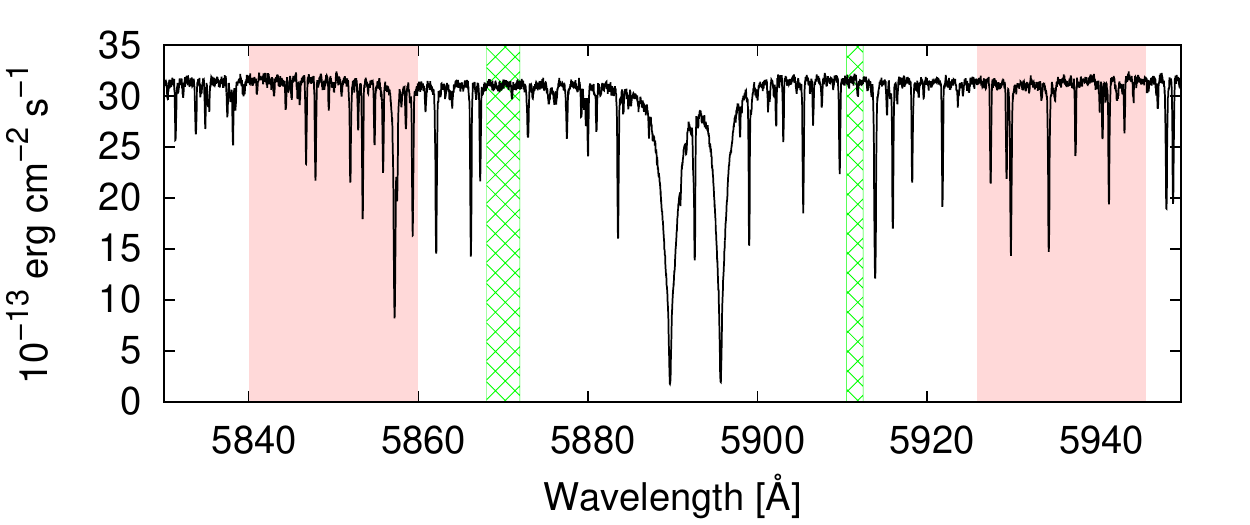}\\
  \includegraphics[width=0.49\textwidth]{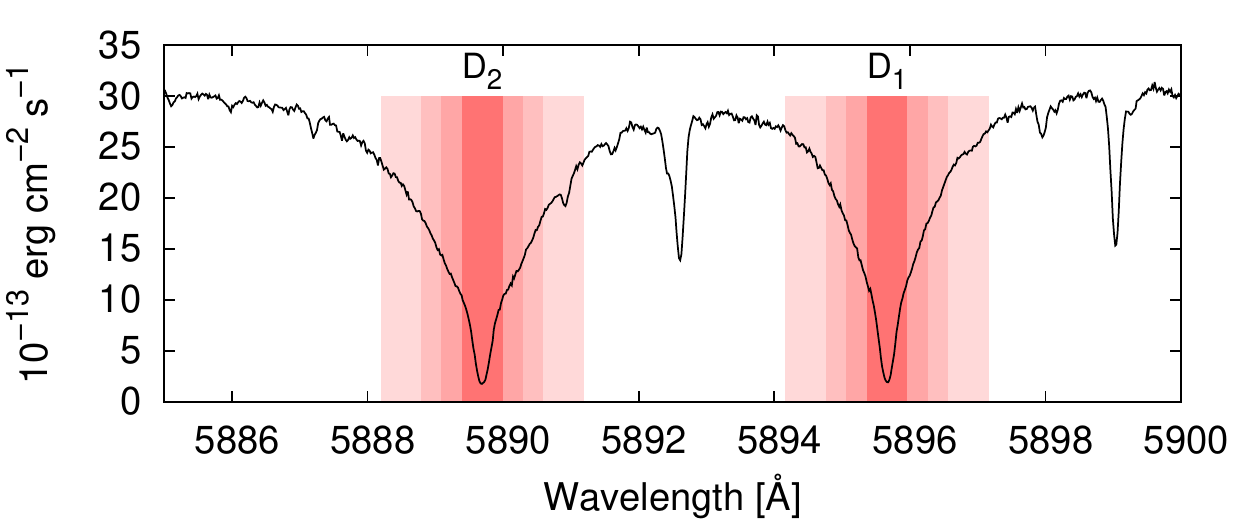}
  \caption{Region covering the \nadd\ lines in our first spectrum.
  Top:  shaded (red) regions show the wide reference bands and  squared
  regions the narrow spectral bands. Bottom: close-up of the \nadd\ lines. The
  shaded (red) regions have half-width of $0.3$, $0.6$, $0.9$, and $1.5$~\AA.
  \label{fig:naspecs}}
\end{figure}

In Fig.~\ref{fig:naspecs}, we show the observed, flux-calibrated spectrum around
the \nadd\ lines. Note that neither a barycentric correction nor any
further normalization was applied to the spectra.
From the observed series of spectra, we extract
feature light curves using
three different feature bands with half-widths of $0.6$~\AA, $0.9$~\AA, and
$1.5$~\AA\ centered on the cores of the sodium lines. These core regions, however,
were disregarded by excluding the inner $\pm 0.3$~\AA\ region.
We then obtained two feature light curves for
both line cores by individually integrating the spectral signal in the blue and red wing.

In our calculations, we
used both wide ($20$~\AA) and narrow reference bands (see
Fig.~\ref{fig:naspecs}). The wide reference bands are defined between
$5840-5860$~\AA\ and $5925.875-5945.875$~\AA. To test the influence of the
reference bands on the outcome, we defined two further pairs of wide reference
bands by shifting both bands toward the \nadd\ lines in two steps of $10$~\AA.
Additionally, we defined narrow reference bands ($5868-5872$~\AA\ and
$5910.5-5912.5$~\AA), which contain no strong spectral lines
(Fig.~\ref{fig:naspecs}).
To correct for the
radial-velocity shift of the individual spectra, we carried out a
cross-correlation with a model spectrum in the $5850-5870$~\AA\ band and
adjusted the start- and end wavelengths of the integration bands according to
the radial-velocity shifts thus derived. With this approach, we account for all present
shifts, whether they are instrumental or physical in origin.

\begin{figure*}[t!]
  \includegraphics[angle=0, width=0.49\textwidth]{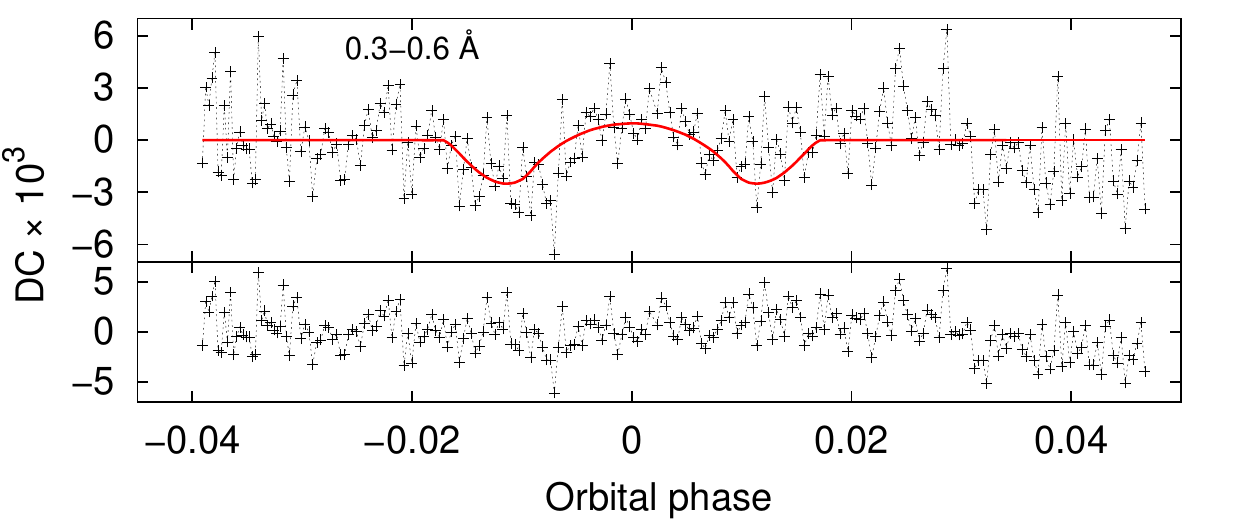}
  \includegraphics[angle=0, width=0.49\textwidth]{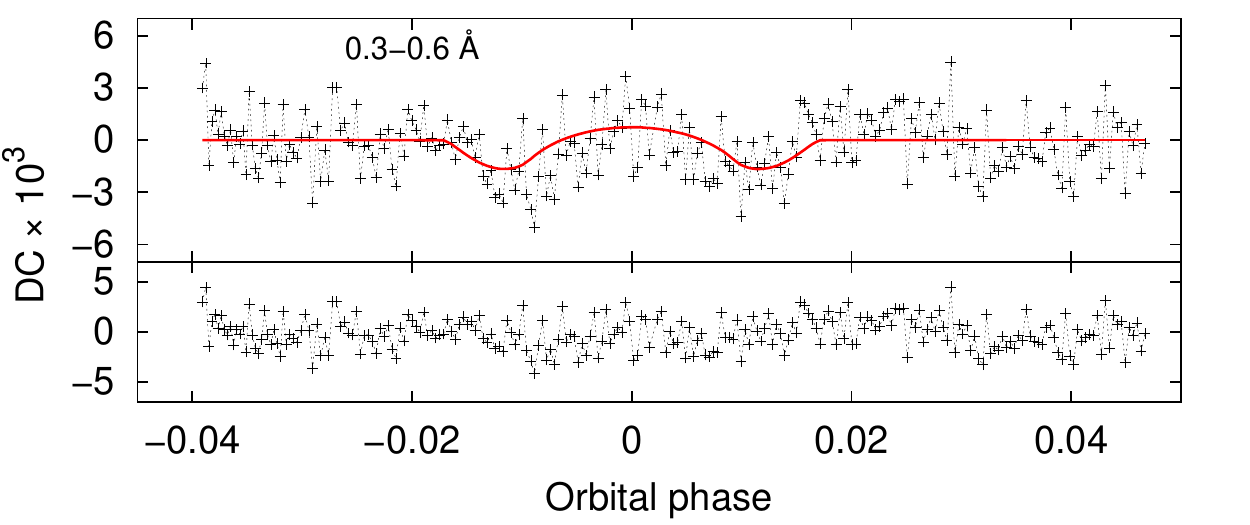}\\
  \includegraphics[angle=0, width=0.49\textwidth]{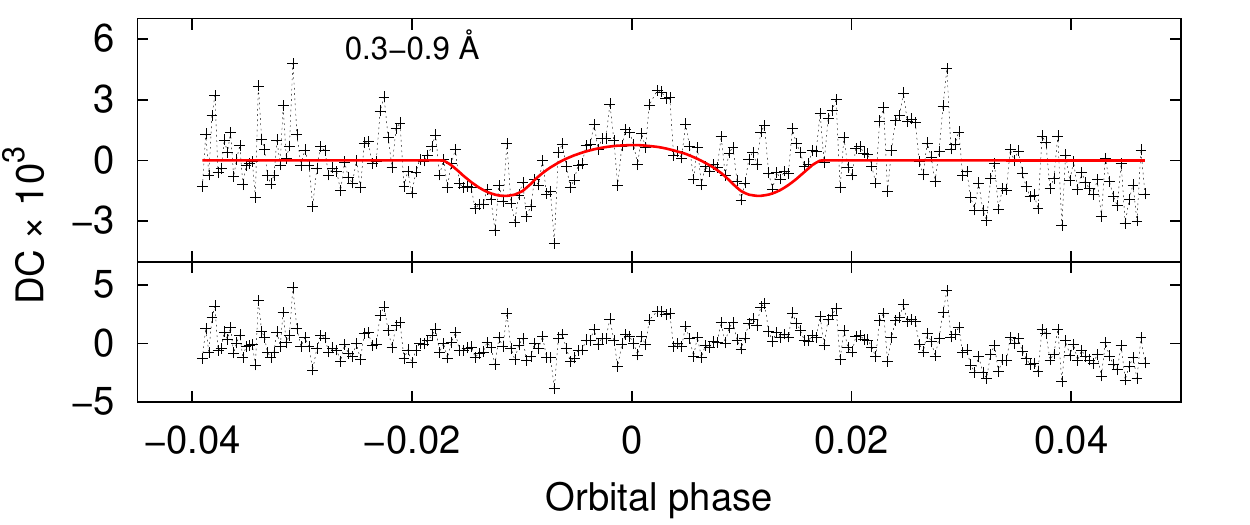}
  \includegraphics[angle=0, width=0.49\textwidth]{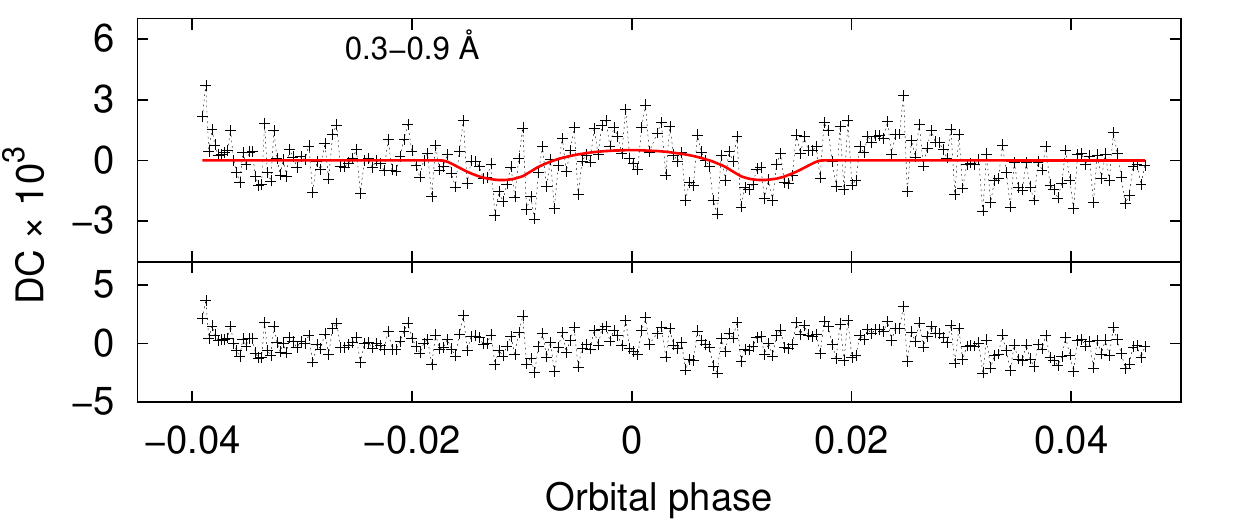}\\
  \includegraphics[angle=0, width=0.49\textwidth]{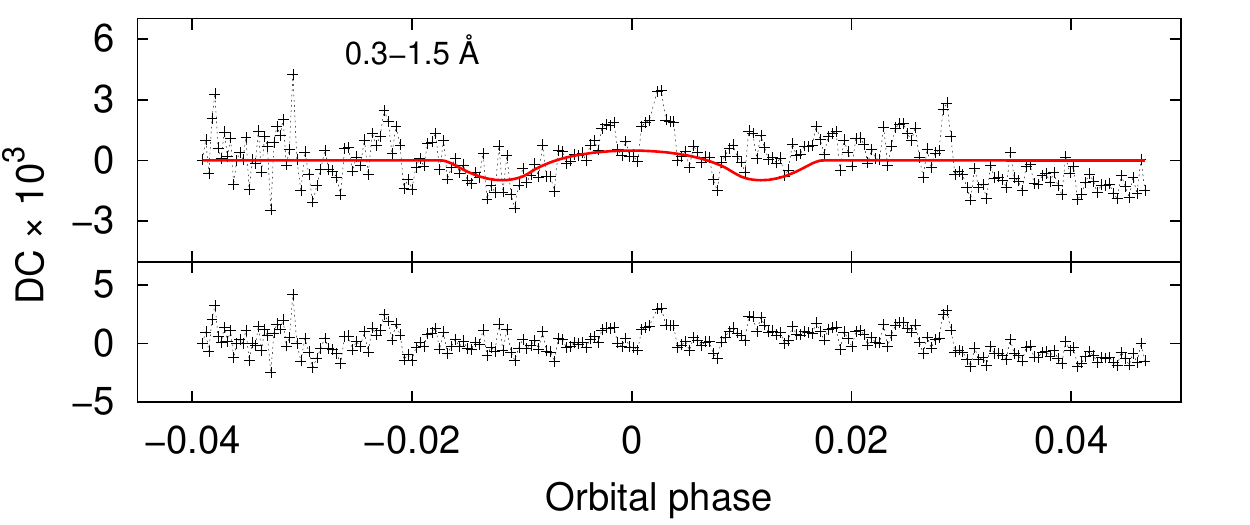}
  \includegraphics[angle=0, width=0.49\textwidth]{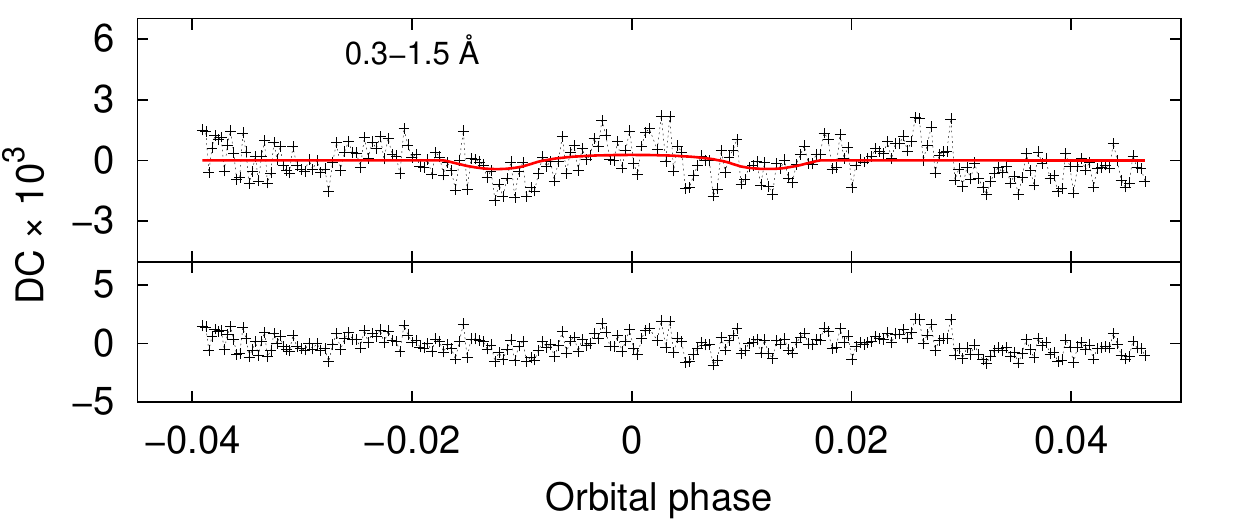}
  \caption{Difference curves for \nal\ $D_2$ (left column) and $D_1$ (right
  column) for three feature bands with half-widths of $0.6$~\AA, $0.9$~\AA, and
  $1.5$~\AA\ (top to bottom). Top panels show the data and the model (solid,
  red line), and the bottom show the residuals. The line core ($\pm 0.3$~\AA\
  around the line center) has been excluded. The reference bands were
  $5840-5860$~\AA\ and $5925.875-5945.875$~\AA. The effect of the CLV is
  different in the two lines and it is weaker for broader spectral bands. 
  \label{fig:NaDCs}}
\end{figure*}

The resulting DCs are shown in
Fig.~\ref{fig:NaDCs} for both line cores
along with model curves
obtained from our simulations of spectral time series.
Around the center of the transit, we find
the behavior of the curves shown in Fig.~\ref{fig:NaDCs} to be consistent with
our model DCs. In particular, the DCs of both sodium lines and all chosen
feature bandwidths show a drop during in- and egress phases and a peak at
transit center. The synthetic curves show that the CLV effect is not identical
for both sodium lines, but a little stronger for the D$_2$ line for the chosen
integration bands. This reflects a difference in width and strength of the
\nadd\ lines, with the D$_2$ being a little broader and deeper in stars like
\hde.

At orbital phase $\approx 0.03$, the observed DCs show a drop for which we
 currently have no definite explanation. We speculate that this drop might be
related to the shift in the curve of the \rme\ seen toward the end of the
observation.
At this point,  the instrumental resolution and radial-velocity shift also show a
general change in their behavior (see Figs.~\ref{fig:tellurRV}). 
Whether the evolution in the DC is ultimately related to an instrumental effect
or a true, physical change in the stellar spectrum, e.g., due to the flare or a
change on the surface caused by a spot remains unknown.

\begin{figure}
  \includegraphics[angle=0, width=0.49\textwidth]{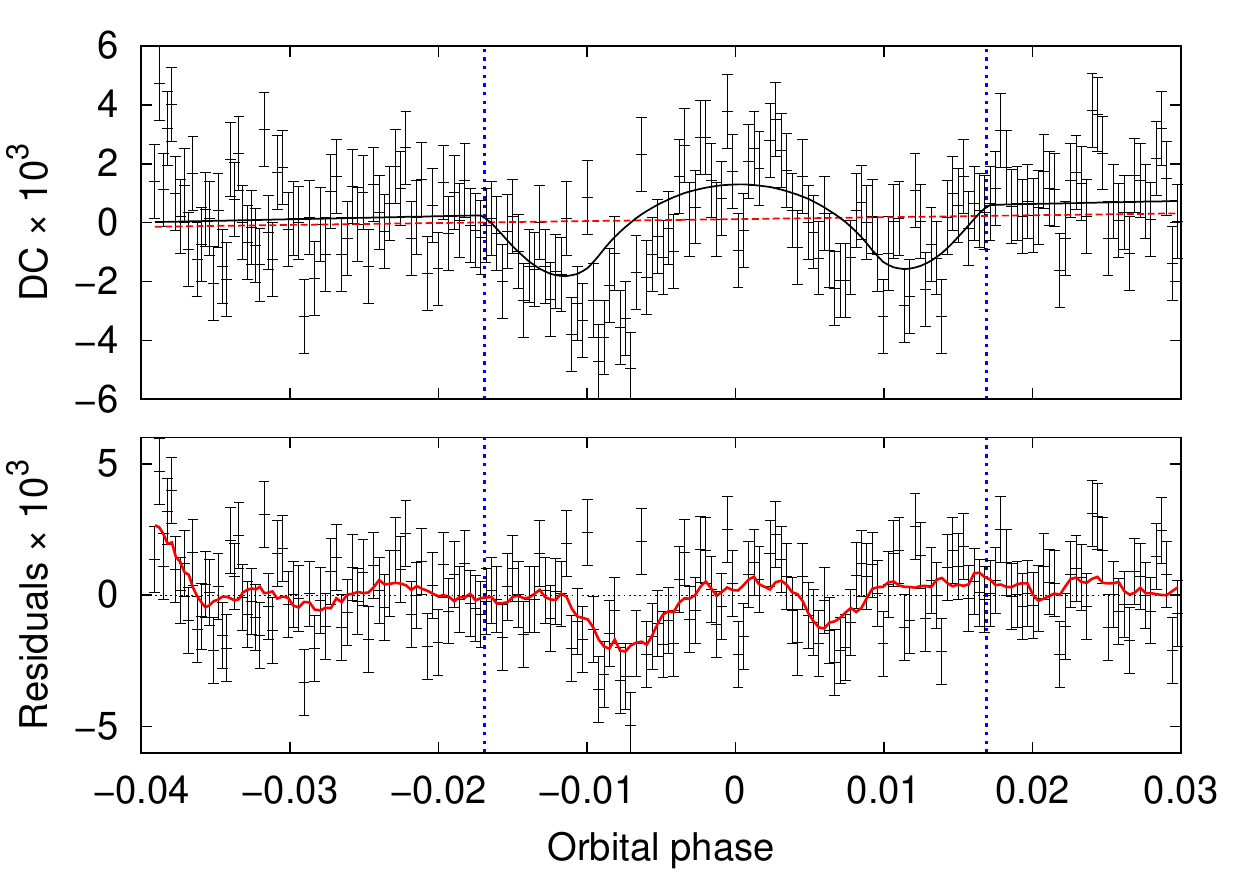}
  \caption{Top: DC observed in the $0.3-0.6$~\AA\ band around
  both \nadd\ cores with respect to narrow reference intervals. The dashed
  (red) curve shows a linear fit and the solid (black) line a linear fit plus
  our synthetic DC model.
  Bottom: residuals with respect to the synthetic model. The solid (red) curve
  shows the residuals smoothed using a running mean with a width of eight~data
  points.
  \label{fig:modelDC}}
\end{figure}

In Fig.~\ref{fig:modelDC}, we show the combined DC from both \nadd\ line cores
($0.3-0.6$~\AA\ band) up to an orbital phase of $0.03$ along with a linear model
and a linear model plus our synthetic DC. The curves were produced using the
narrow reference band. However, we find little change if other reference bands
are used. The error was estimated as the scatter in the observed DC. Clearly,
the synthetic model DC explains the general behavior of the observation. In
particular, we obtain a $\chi^2$ value of $369$ using only the linear model and
a value of $279$ if the full model is used. In both cases, only the two
parameters of the linear model were fitted. The profile of the synthetic DC is
completely determined by the transit geometry and the stellar model
atmosphere. Therefore, we are left with $195$ degrees of freedom. It seems, however,
that the observed DC shows further structure not accounted for by the model.
Before and after mid-transit time, there are two drops in the residuals. The
structure is, however, not entirely symmetric with the deviation stronger
during the first half of the transit.

To study the effect of variable telluric absorption on our measurements, we
used the \texttt{molecfit} code to correct our spectra for telluric
contamination \citep[][]{Smette2015, Kausch2015}. The DCs
produced using the spectra corrected for telluric absorption do
not significantly differ from those that are not corrected for telluric absorption. The main difference is a systematic
pattern changing in phase with air mass. The amplitude of this pattern is,
however, only $0.5\times 10^{-3}$. Formally, we obtained a reduced $\chi^2$
value of 1.48 for the CLV model with telluric correction, which is even a little
higher than the value of 1.43 obtained without telluric correction. As the CLV-related wave-like
pattern remains essentially unaffected, we conclude that variable telluric
absorption is unlikely to be the origin of the observed behavior.

Finally, we verified that the change in instrumental resolution detected in the
UVES spectra (see Sect.~\ref{sec:instrResol}) is not responsible for the observed
behavior.
In particular, we simulated DCs applying instrumental resolutions of $55\,000$ and
$65\,000$ representing the observed change in resolution (see
Sect.~\ref{sec:instrResol}).
The resulting change in the DCs remains too small to be of any 
relevance in the current study.

\section{Summary and discussion}
\label{sec:SummAndDis}

We present DCs of the \cahk\ and \nadd\
lines derived from transit spectroscopy of \hde\ obtained with UVES. 
Our DCs clearly show the effect of differential CLV in the wings of the
\cahk\ and \nadd\ lines of \hde. 

\subsection{The center-to-limb variation}
Our analysis of the \cahk\ line cores revealed strong variability. Most notably,
an increase in the strength of the cores, which we interpret as a stellar flare,
was observed at about mid-transit time. We found that the core DC
obtained before the flare is better reproduced by a model 
assuming no limb darkening in the \cahk\
line cores than by a model with photospheric limb darkening. This finding
is compatible with a homogeneous brightness of the chromospheric core emission
across the stellar disk. Yet, the actual spatial distribution is
probably more complex \citep[][]{Llama2015}.
The \cahk\ emission line cores also show
variability before the ingress, which is probably unrelated to the
transit. While an
earlier ingress either due to planetary material or an extended stellar
chromosphere is conceivable \citep[e.g.,][]{Czesla2012, Haswell2012},
intrinsic variability, independent of the transit, appears to be
a more plausible interpretation  in  view of the flare observed later.
In any case, it remains difficult
to distinguish between effects caused by the transiting planetary disk and
intrinsic stellar variability in our data set. 

To limit the impact of activity, we 
focused our analysis on
the wings of the Fraunhofer \cahk\ and \nadd\ lines. These are expected to be
less strongly influenced by activity, but are still sensitive to the stellar
CLV.
The shapes of both the DCs based on the \cahk\ and \nadd\ line wings clearly
show the behavior expected from the stellar CLV. In particular, we find the line
wings to be limb brightened compared to the reference continuum.

The observed DCs are well reproduced by a model composed
of a synthetic DC and a low-order polynomial. Including the synthetic DC model
clearly improves the fit compared to a low-order polynomial alone. In
particular, the reduced $\chi^2$ values changed from $1.31$ to $1.02$ in
for the \cahk\ lines and from $1.89$ to $1.43$ for the \nadd\ lines.
Nonetheless, we still obtain some correlated residuals, 
especially in the
modeling of the DC derived from the \nadd\ lines. 

There may be several reasons for
these residuals: first, the modeling may be incomplete.
\citet{Hayek2012} find better fits to transit light curves with CLV models based
on three-dimensional LTE model atmospheres, but our modeling is based on
one-dimensional LTE models. Also, non-LTE effects are expected to
influence the formation of the strong Fraunhofer lines
\citep[e.g.,][]{Lind2011}.
Because the orbit of \hde\,b shows only minor misalignment
with the stellar spin axis
\citep[$\beta=-\lambda=0.85^{+0.28}_{-0.32}$ $^{\circ}$,][]{Triaud2009},
we expect that this kind of an effect should be symmetric with respect to mid-transit
time.
Therefore, atmospheric model deficiencies can hardly account for all the
residuals.
Second, there may be instrumental effects. Our data show a complex pattern of
instrumental radial velocity shift, and we identified residuals, which we
attributed to instrumental effects, in the difference spectra of the \cahk\
lines. Clearly, instrumental effects could give rise to residuals on the
observed scale.
Between orbital phase $-0.01$
and $0$, there seems to be a structure in the residuals of both the \cahk\ and
\nadd\ lines, which may have a physical origin.
Stellar activity provides viable candidates for this kind of structure.

\subsection{The influence of stellar activity on the difference curve}
Starspots and flares are prominent manifestations of stellar activity.
In the following, we briefly discuss their potential impact
on the observed DCs.
Indeed, starspot
occultations have been observed in transit light curves of
\hde\,b earlier \citep[e.g.,][]{Pont2008, Sing2011}, which makes
a spot crossing event a plausible scenario. In particular,
as we certainly observed \hde\ in an active phase as evidenced by the tentative
flare starting near mid-transit time.

Starspots crossed by the transiting planet cause ``bumps'' in the broadband
transit light curve because comparably darker regions of the stellar disk are
occulted \citep[e.g.,][]{Pont2008, Czesla2009}.
For the DC, the height difference of the bump in the feature and
reference light curve
is relevant. As the wings of the \nal\ lines become deeper for lower effective
temperatures (see Fig.~\ref{fig:Na}), the spot appears darker in the wings of
the \nal\ lines than in the surrounding continuum,
given that its spectrum can still be
described by an otherwise normal photospheric model. This gives rise to a
more pronounced bump in feature light curves covering the \nal\ line wings and,
thus, a net increase in the DC. The same applies to the \cahk\ lines.
The amplitude of the effect depends
on the spectral characteristics and shape of the spot. For a large spot with
effective temperature of $4000$~K \citep[][]{Pont2008}, we expect that the
spot-induced bump can be of the same scale as the amplitude of the actual DC
because the change in the profile of the \nal\ lines is quite pronounced between
$4000$~K and $5000$~K (see Fig.~\ref{fig:Na}).

Also starspots not occulted by
the planet affect the DC because they cause an effective wavelength-dependent
change in the planet-to-star radius ratio and, thus, alter the normalization
of the light curves, which changes the transit depth
\citep[][]{Sing2008LD, Czesla2009}. While the effect of the starspot crossing is
local, unocculted spots modify the entire transit light curve,
given that they do not rotate on or off the disk. At least in the case of \hde,
this seems improbable, because we only observed  a little more than one percent
of the stellar rotation period.

The flare observed as a rise in the \cahk\ emission line cores at about
mid-transit time could influence the DC both via chromospheric and photospheric
emission. During flares, the strength of chromospheric emission lines including
\cahk\ and \nadd\, are expected to increase \citep[e.g.,][]{Fuhrmeister2004}.
While we exclude the cores in the analysis of the line wings, some contribution
could still be present. 
A fraction of the chromosphere and upper photosphere is thought to be heated to
temperatures of $\approx 10\,000$~K \citep{Haisch1991}.
Effectively, a part of the quiescent photospheric emission would be
replaced by emission from a $10\,000$~K hot spot. The spectra of hot
photospheres are dominated by broad Balmer lines, while lines of other ions
(including \nal) tend to be narrow. While the exact impact of the flare depends
on its location, size, and spectrum, its effect should be temporally limited,
and we expect the main contribution to take place during the leading impulsive phase.

\subsection{The planetary atmosphere of \hde\,b seen in \nadd}
Difference curves have been used to detect planetary excess absorption in the
\nadd\ lines during transit.
While there are residuals in our DC not accounted for by our modeling,
their temporal behavior is rather untypical for a signal related to the planetary
atmosphere (see Fig.~\ref{fig:modelDC}). In particular, only limited
sections before and after mid-transit time show deviations, while the planetary
atmosphere should be visible during the entire transit.
The presence of residuals of ultimately unknown origin makes it difficult to
derive a reliable estimate of the atmospheric excess
absorption in the wings of the \nadd\ lines in the frame of our current
analysis.

Detections of in-transit excess absorption in the \nadd\ lines attributable to
the planetary atmosphere have been presented by \citet{Redfield2008},
\citet{Huitson2012}, and \citet{Wyttenbach2015}. Both \citet{Huitson2012} and
\citet{Wyttenbach2015} find the bulk of planetary atmospheric absorption
concentrated in the cores of the \nadd\ lines. \citeauthor{Wyttenbach2015}
derived a FWHM of $0.52\pm0.08$~\AA\ for their
signal.
As we exclude the inner $\pm
0.3$~\AA\ region around the \nadd\ line cores in our analysis to minimize
contamination by stellar activity, we also exclude about $80$\,\% of their
signal. In the \nadd\ lines, they report a relative transit depth of
$(3.2\pm0.31)\times 10^{-3}$ so that the maximum signal we expect is about
$0.6\times 10^{-3}$. Therefore, we do not regard the lack of an obvious signal related to the
planetary atmosphere in our current analysis to be a contradiction of  the findings of
\citet{Wyttenbach2015}. We also do not exclude a contribution by the planetary
atmosphere in the DC.

An inspection of Fig.~3 in \citet{Wyttenbach2015} suggests that their curves
may also show the effect of the stellar CLV. The curves seem to show a bump in
the middle of the transit, which resembles the DCs we obtain for limb-brightened
features. In fact, our simulation for their $12$~\AA\ wide band indicate a
peak-to-peak amplitude of about $0.06$\,\%
for the DC, which seems to match the distribution of their residuals. In the
smaller $0.75$~\AA\ band, the expected CLV-induced variation is on the order of $0.4$\,\% (see
Fig.~\ref{fig:NaDCs}), which is again compatible with their findings. We
emphasize, however, that their detection of excess sodium absorption would not
be challenged by this. From our models, we obtain a net CLV-induced transit
depth of $0.006$\,\% in the $12$~\AA\ wide band, which is on the order of their
error. As this shift would be systematic, the level of sodium absorption derived
by \citet{Wyttenbach2015} may be underestimated by this rather small
amount.

\subsection{The impact of the temporal sampling of the transit}

The spectrum of the transited stellar disk evolves
continuously as the planet proceeds along its path and therefore, the DC is
time dependent. This implies that the temporal sampling of the transit is
highly relevant in the analysis of transit spectroscopy.
If, for instance, only
the central part of the transit of \hde\,b was covered, the wings of the
\nadd\ and \cahk\ lines would actually show net emission. The reverse is true,
if only the in- and egress phases were covered. This effect occurs without
requiring any change in the physical radius of the planet.

\subsection{Relevance of the CLV in large planets}
\label{sec:smallPlanet}
If the planetary atmosphere can be treated as a thin shell enshrouding
the planet, it extends the planetary disk seen during transit by a small amount,
$\Delta R_p$. The area, $A_a$, of the resulting atmospheric annulus reads
\begin{equation}
  \frac{A_a}{\pi R_s^2} = \frac{\pi(R_p+\Delta R_p)^2 - \pi R_s^2}{\pi R_s^2}
  \approx 2 \left(\frac{R_p}{R_s}\right)\left(\frac{\Delta R_p}{R_s}\right) =
  2p\Delta p \; .
\end{equation}
When the strength of the atmospheric absorption is proportional to
the atmospheric area, the effect observable in the spectrum is proportional to
the planet-to-star radius ratio, $p$, and the extent of its
atmosphere. Therefore, large planet-to-star radius ratios and large atmospheric
scale heights are favorable to detect atmospheric absorption. 

A simplified model of the transit light curve with quadratic
limb darkening can be obtained for small planets, for which the
normalized transit light curve as a function of the limb angle, $\mu$,
reads
\begin{equation}
  n_{sp}(a, b, \mu, p) = 1 - p^2 \left( \frac{1 - a(1-\mu) -
  b(1-\mu)^2}{1-\frac{a}{3}-\frac{b}{6}}\right) \; .
  \label{eq:tran}
\end{equation}
Here, the limb darkening is parameterized according to Eq.~\ref{eq:qld}
\citep[cf.][Eq.~B.1]{Mueller2013}.
Accordingly, the DC can be approximated by
\begin{equation}
  DC_{sp}(\mu, p) = n_{sp}(a_f, b_f, \mu, p) - n_{sp}(a_r, b_r, \mu, p)
  \sim p^2 \; ,
  \label{eq:dcapprox}
\end{equation}
which depends on the properties of the normalized feature (f) and reference (r)
light curves.
While this simplified model does not reproduce the
in- and egress phases appropriately,
Eq.~\ref{eq:dcapprox} can be used to estimate the value of the DC in the
transit center. For the cases discussed in Sect.~\ref{sec:diffandrat}, we obtain
values of $-8.9\times 10^{-4}$ for the limb-darkened feature and $+10.4\times
10^{-4}$ for the limb-brightened feature, which both agree well with the results
obtained from the more elaborate models (cf. Fig.~\ref{fig:ldlb}).

Equation~\ref{eq:dcapprox} demonstrates that the amplitude of the
CLV-induced effect on the DC is proportional to the squared planet-to-star
radius ratio, while the strength of the signal induced by the planetary
atmosphere grows linearly with the radius ratio. Additionally, smaller stars,
which potentially provide more favorable (i.e., larger) radius ratios also tend
to be cooler. In cool stars, however, the CLV-induced variation is also more
pronounced. This underlines the importance of an accurate knowledge and modeling
of the stellar CLV in the study of planetary atmospheres. 

\section{Conclusion}
\label{sec:Conclusion}
We clearly detect the CLV-induced effect in a spectral transit time series of
\hde, and our findings are generally compatible with predictions from 1D-LTE
models.
Remaining deviations may be explained by more elaborate modeling of the stellar
CLV, stellar activity, instrumental effects, or a combination thereof. The
CLV-induced effect is stronger for cooler stars and its relative importance
increases for larger planet-to-star radius ratios. Our results demonstrate that
the variation in the stellar CLV across individual spectral lines can be
detected with transit spectroscopy and that the effect should be taken into
account in the study of planetary atmospheres via transmission spectroscopy.

Transiting planets provide a valuable opportunity to resolve the stellar
disk, which has been used to study the surface distribution of
starspots \citep[e.g.,][]{Pont2008, Huber2010, SanchisOjeda2011}.
Here we show that the moving planetary disk can also be used to
study the spatially resolved stellar spectrum, which is required to constrain
three-dimensional hydrodynamic models of the stellar photosphere
\citep{Dravins2014}.
 
Given appropriate modeling, the prominent CLV in the stellar Fraunhofer lines
could also be used to reconstruct the planetary path across the stellar disk in
terms of the temporal change in the occulted limb angle. In its use of the
spectrum, this approach resembles the \rme, which is frequently used to study
the orbit geometry of exoplanets \citep[e.g.,][]{Winn2006, Triaud2009}. While
the strength of this effect depends on the properties of the stellar
atmosphere, stellar rotation is not required to see it.
Clearly,  the CLV in the specific ensemble of stellar spectral lines used to
derive the radial velocity shifts should also be taken into account in the
modeling of the \rme. We speculate that the correlated residuals in the \rmc\
curve obtained by \citet{Winn2006}, \citet{Triaud2009}, and in this work could be attributable to the stellar CLV.

As noted by \citet{Yan2015}, the CLV-induced effect on the light curves derived
from transit spectroscopy becomes more pronounced when narrow spectral bands
are used for integration in strong spectral lines. Although the use of broader
bands reduces the influence of the CLV, a number of studies have found the
planetary atmospheric signal to be concentrated in the cores of the line.
Therefore, the use of broader bands may not always be desirable. At any rate, an
accurate treatment of the stellar CLV is indispensable in the
analysis of transit spectroscopy of exoplanets.

\begin{acknowledgements}
TK acknowledges support by the DFG program \mbox{CZ 222/1-1}. TK and SK
acknowledge support from the RTG~1351 (``Extrasolar planets and their host
stars''). UW acknowledges support from the DLR
under grant 50OR0105.

\end{acknowledgements}

\bibliographystyle{aa}
\bibliography{ulun.bib}

\begin{thebibliography}{73}
\expandafter\ifx\csname natexlab\endcsname\relax\def\natexlab#1{#1}\fi

\bibitem[{{Agol} {et~al.}(2010){Agol}, {Cowan}, {Knutson}, {Deming}, {Steffen},
  {Henry}, \& {Charbonneau}}]{Agol2010}
{Agol}, E., {Cowan}, N.~B., {Knutson}, H.~A., {et~al.} 2010, \apj, 721, 1861

\bibitem[{{Athay} {et~al.}(1972){Athay}, {Lites}, {White}, \&
  {Brault}}]{Athay1972}
{Athay}, R.~G., {Lites}, B.~W., {White}, O.~R., \& {Brault}, J.~W. 1972,
  \solphys, 24, 18

\bibitem[{{Baliunas} {et~al.}(1995){Baliunas}, {Donahue}, {Soon}, {Horne},
  {Frazer}, {Woodard-Eklund}, {Bradford}, {Rao}, {Wilson}, {Zhang}, {Bennett},
  {Briggs}, {Carroll}, {Duncan}, {Figueroa}, {Lanning}, {Misch}, {Mueller},
  {Noyes}, {Poppe}, {Porter}, {Robinson}, {Russell}, {Shelton}, {Soyumer},
  {Vaughan}, \& {Whitney}}]{Baliunas1995}
{Baliunas}, S.~L., {Donahue}, R.~A., {Soon}, W.~H., {et~al.} 1995, \apj, 438,
  269

\bibitem[{{Balthasar} {et~al.}(1982){Balthasar}, {Thiele}, \&
  {Woehl}}]{Balthasar1982}
{Balthasar}, H., {Thiele}, U., \& {Woehl}, H. 1982, \aap, 114, 357

\bibitem[{{Bouchy} {et~al.}(2005){Bouchy}, {Udry}, {Mayor}, {Moutou}, {Pont},
  {Iribarne}, {da Silva}, {Ilovaisky}, {Queloz}, {Santos}, {S{\'e}gransan}, \&
  {Zucker}}]{Bouchy2005}
{Bouchy}, F., {Udry}, S., {Mayor}, M., {et~al.} 2005, \aap, 444, L15

\bibitem[{{Brown}(2001)}]{Brown2001}
{Brown}, T.~M. 2001, \apj, 553, 1006

\bibitem[{{Burton} {et~al.}(2015){Burton}, {Watson}, {Rodr{\'{\i}}guez-Gil},
  {Skillen}, {Littlefair}, {Dhillon}, \& {Pollacco}}]{Burton2015}
{Burton}, J.~R., {Watson}, C.~A., {Rodr{\'{\i}}guez-Gil}, P., {et~al.} 2015,
  \mnras, 446, 1071

\bibitem[{{Caccin} {et~al.}(1985){Caccin}, {Cavallini}, {Ceppatelli},
  {Righini}, \& {Sambuco}}]{Caccin1985}
{Caccin}, B., {Cavallini}, F., {Ceppatelli}, G., {Righini}, A., \& {Sambuco},
  A.~M. 1985, \aap, 149, 357

\bibitem[{{Castelli} \& {Kurucz}(2004)}]{Castelli2004}
{Castelli}, F. \& {Kurucz}, R.~L. 2004, ArXiv Astrophysics e-prints
  [\eprint{arXiv:astro-ph/0405087}]

\bibitem[{{Charbonneau} {et~al.}(2000){Charbonneau}, {Brown}, {Latham}, \&
  {Mayor}}]{Charbonneau2000}
{Charbonneau}, D., {Brown}, T.~M., {Latham}, D.~W., \& {Mayor}, M. 2000, \apjl,
  529, L45

\bibitem[{{Charbonneau} {et~al.}(2002){Charbonneau}, {Brown}, {Noyes}, \&
  {Gilliland}}]{Charbonneau2002}
{Charbonneau}, D., {Brown}, T.~M., {Noyes}, R.~W., \& {Gilliland}, R.~L. 2002,
  \apj, 568, 377

\bibitem[{{Claret}(2004)}]{Claret2004}
{Claret}, A. 2004, \aap, 428, 1001

\bibitem[{{Clough} {et~al.}(2005){Clough}, {Shephard}, {Mlawer}, {Delamere},
  {Iacono}, {Cady-Pereira}, {Boukabara}, \& {Brown}}]{Clough2005}
{Clough}, S.~A., {Shephard}, M.~W., {Mlawer}, E.~J., {et~al.} 2005, \jqsrt, 91,
  233

\bibitem[{{Croll} {et~al.}(2007){Croll}, {Matthews}, {Rowe}, {Gladman},
  {Miller-Ricci}, {Sasselov}, {Walker}, {Kuschnig}, {Lin}, {Guenther},
  {Moffat}, {Rucinski}, \& {Weiss}}]{Croll2007}
{Croll}, B., {Matthews}, J.~M., {Rowe}, J.~F., {et~al.} 2007, \apj, 671, 2129

\bibitem[{{Czesla} {et~al.}(2009){Czesla}, {Huber}, {Wolter}, {Schr{\"o}ter},
  \& {Schmitt}}]{Czesla2009}
{Czesla}, S., {Huber}, K.~F., {Wolter}, U., {Schr{\"o}ter}, S., \& {Schmitt},
  J.~H.~M.~M. 2009, \aap, 505, 1277

\bibitem[{{Czesla} {et~al.}(2012){Czesla}, {Schr{\"o}ter}, {Wolter}, {von
  Essen}, {Huber}, {Schmitt}, {Reichart}, \& {Moore}}]{Czesla2012}
{Czesla}, S., {Schr{\"o}ter}, S., {Wolter}, U., {et~al.} 2012, \aap, 539, A150

\bibitem[{{Dravins} {et~al.}(2014){Dravins}, {Ludwig}, {Dahl{\'e}n}, \&
  {Pazira}}]{Dravins2014}
{Dravins}, D., {Ludwig}, H.-G., {Dahl{\'e}n}, E., \& {Pazira}, H. 2014, ArXiv
  e-prints [\eprint[arXiv]{1408.1402}]

\bibitem[{{Fang} {et~al.}(1992){Fang}, {Hiei}, {Yin}, \& {Gan}}]{Fang1992}
{Fang}, C., {Hiei}, E., {Yin}, S.-Y., \& {Gan}, W.-Q. 1992, \pasj, 44, 63

\bibitem[{{Fossati} {et~al.}(2010){Fossati}, {Haswell}, {Froning}, {Hebb},
  {Holmes}, {Kolb}, {Helling}, {Carter}, {Wheatley}, {Collier Cameron},
  {Loeillet}, {Pollacco}, {Street}, {Stempels}, {Simpson}, {Udry}, {Joshi},
  {West}, {Skillen}, \& {Wilson}}]{Fossati2010}
{Fossati}, L., {Haswell}, C.~A., {Froning}, C.~S., {et~al.} 2010, \apjl, 714,
  L222

\bibitem[{{Fuhrmeister} {et~al.}(2008){Fuhrmeister}, {Liefke}, {Schmitt}, \&
  {Reiners}}]{Fuhrmeister2008}
{Fuhrmeister}, B., {Liefke}, C., {Schmitt}, J.~H.~M.~M., \& {Reiners}, A. 2008,
  \aap, 487, 293

\bibitem[{{Fuhrmeister} \& {Schmitt}(2004)}]{Fuhrmeister2004}
{Fuhrmeister}, B. \& {Schmitt}, J.~H.~M.~M. 2004, \aap, 420, 1079

\bibitem[{Gray(2008)}]{Gray2008}
Gray, D. 2008, The Observation and Analysis of Stellar Photospheres, 3rd edn.
  (Cambridge University Press)

\bibitem[{{Gray} \& {Brown}(2006)}]{Gray2006}
{Gray}, D.~F. \& {Brown}, K.~I.~T. 2006, \pasp, 118, 399

\bibitem[{{Gray} \& {Corbally}(1994)}]{Gray1994}
{Gray}, R.~O. \& {Corbally}, C.~J. 1994, \aj, 107, 742

\bibitem[{{Grossmann-Doerth}(1994)}]{Grossmann-Doerth1994}
{Grossmann-Doerth}, U. 1994, \aap, 285, 1012

\bibitem[{{Haisch} {et~al.}(1991){Haisch}, {Strong}, \& {Rodono}}]{Haisch1991}
{Haisch}, B., {Strong}, K.~T., \& {Rodono}, M. 1991, \araa, 29, 275

\bibitem[{{Haswell} {et~al.}(2012){Haswell}, {Fossati}, {Ayres}, {France},
  {Froning}, {Holmes}, {Kolb}, {Busuttil}, {Street}, {Hebb}, {Collier Cameron},
  {Enoch}, {Burwitz}, {Rodriguez}, {West}, {Pollacco}, {Wheatley}, \&
  {Carter}}]{Haswell2012}
{Haswell}, C.~A., {Fossati}, L., {Ayres}, T., {et~al.} 2012, \apj, 760, 79

\bibitem[{{Hayek} {et~al.}(2012){Hayek}, {Sing}, {Pont}, \&
  {Asplund}}]{Hayek2012}
{Hayek}, W., {Sing}, D., {Pont}, F., \& {Asplund}, M. 2012, \aap, 539, A102

\bibitem[{{Huber} {et~al.}(2010){Huber}, {Czesla}, {Wolter}, \&
  {Schmitt}}]{Huber2010}
{Huber}, K.~F., {Czesla}, S., {Wolter}, U., \& {Schmitt}, J.~H.~M.~M. 2010,
  \aap, 514, A39

\bibitem[{{Huitson} {et~al.}(2012){Huitson}, {Sing}, {Vidal-Madjar},
  {Ballester}, {Lecavelier des Etangs}, {D{\'e}sert}, \& {Pont}}]{Huitson2012}
{Huitson}, C.~M., {Sing}, D.~K., {Vidal-Madjar}, A., {et~al.} 2012, \mnras,
  422, 2477

\bibitem[{{Jensen} {et~al.}(2011){Jensen}, {Redfield}, {Endl}, {Cochran},
  {Koesterke}, \& {Barman}}]{Jensen2011}
{Jensen}, A.~G., {Redfield}, S., {Endl}, M., {et~al.} 2011, \apj, 743, 203

\bibitem[{{Kausch} {et~al.}(2015){Kausch}, {Smette}, {Kimeswenger}, {Barden},
  {Szyszka}, {Jones}, {Sana}, {Horst}, \& {Kerber}}]{Kausch2015}
{Kausch}, W., {Smette}, S.~N.~A., {Kimeswenger}, S., {et~al.} 2015, ArXiv
  e-prints [\eprint[arXiv]{1501.07265}]

\bibitem[{{Knutson} {et~al.}(2010){Knutson}, {Howard}, \&
  {Isaacson}}]{Knutson2010}
{Knutson}, H.~A., {Howard}, A.~W., \& {Isaacson}, H. 2010, \apj, 720, 1569

\bibitem[{{Kurucz}(1970)}]{Kurucz1970}
{Kurucz}, R.~L. 1970, SAO Special Report, 309

\bibitem[{{Lalitha} {et~al.}(2013){Lalitha}, {Fuhrmeister}, {Wolter},
  {Schmitt}, {Engels}, \& {Wieringa}}]{Lalitha2013}
{Lalitha}, S., {Fuhrmeister}, B., {Wolter}, U., {et~al.} 2013, \aap, 560, A69

\bibitem[{{Lecavelier Des Etangs} {et~al.}(2010){Lecavelier Des Etangs},
  {Ehrenreich}, {Vidal-Madjar}, {Ballester}, {D{\'e}sert}, {Ferlet},
  {H{\'e}brard}, {Sing}, {Tchakoumegni}, \& {Udry}}]{Lecavelier2010}
{Lecavelier Des Etangs}, A., {Ehrenreich}, D., {Vidal-Madjar}, A., {et~al.}
  2010, \aap, 514, A72

\bibitem[{{Lind} {et~al.}(2011){Lind}, {Asplund}, {Barklem}, \&
  {Belyaev}}]{Lind2011}
{Lind}, K., {Asplund}, M., {Barklem}, P.~S., \& {Belyaev}, A.~K. 2011, \aap,
  528, A103

\bibitem[{{Linsky} {et~al.}(2010){Linsky}, {Yang}, {France}, {Froning},
  {Green}, {Stocke}, \& {Osterman}}]{Linsky2010}
{Linsky}, J.~L., {Yang}, H., {France}, K., {et~al.} 2010, \apj, 717, 1291

\bibitem[{{Llama} \& {Shkolnik}(2015)}]{Llama2015}
{Llama}, J. \& {Shkolnik}, E.~L. 2015, \apj, 802, 41

\bibitem[{{Mandel} \& {Agol}(2002)}]{Mandel2002}
{Mandel}, K. \& {Agol}, E. 2002, \apjl, 580, L171

\bibitem[{{Mayor} \& {Queloz}(1995)}]{Mayor1995}
{Mayor}, M. \& {Queloz}, D. 1995, \nat, 378, 355

\bibitem[{{McLaughlin}(1924)}]{McLaughlin1924}
{McLaughlin}, D.~B. 1924, \apj, 60, 22

\bibitem[{{Melo} {et~al.}(2006){Melo}, {Santos}, {Pont}, {Guillot},
  {Israelian}, {Mayor}, {Queloz}, \& {Udry}}]{Melo2006}
{Melo}, C., {Santos}, N.~C., {Pont}, F., {et~al.} 2006, \aap, 460, 251

\bibitem[{{M{\"u}ller} {et~al.}(2013){M{\"u}ller}, {Huber}, {Czesla}, {Wolter},
  \& {Schmitt}}]{Mueller2013}
{M{\"u}ller}, H.~M., {Huber}, K.~F., {Czesla}, S., {Wolter}, U., \& {Schmitt},
  J.~H.~M.~M. 2013, \aap, 560, A112

\bibitem[{{Narita} {et~al.}(2005){Narita}, {Suto}, {Winn}, {Turner}, {Aoki},
  {Leigh}, {Sato}, {Tamura}, \& {Yamada}}]{Narita2005}
{Narita}, N., {Suto}, Y., {Winn}, J.~N., {et~al.} 2005, \pasj, 57, 471

\bibitem[{{Neckel} \& {Labs}(1994)}]{Neckel1994}
{Neckel}, H. \& {Labs}, D. 1994, \solphys, 153, 91

\bibitem[{{Ohta} {et~al.}(2005){Ohta}, {Taruya}, \& {Suto}}]{Ohta2005}
{Ohta}, Y., {Taruya}, A., \& {Suto}, Y. 2005, \apj, 622, 1118

\bibitem[{{Pierce} \& {Slaughter}(1977)}]{Pierce1977}
{Pierce}, A.~K. \& {Slaughter}, C.~D. 1977, \solphys, 51, 25

\bibitem[{{Pont} {et~al.}(2008){Pont}, {Knutson}, {Gilliland}, {Moutou}, \&
  {Charbonneau}}]{Pont2008}
{Pont}, F., {Knutson}, H., {Gilliland}, R.~L., {Moutou}, C., \& {Charbonneau},
  D. 2008, \mnras, 385, 109

\bibitem[{{Poppenhaeger} {et~al.}(2013){Poppenhaeger}, {Schmitt}, \&
  {Wolk}}]{Poppenhaeger2013}
{Poppenhaeger}, K., {Schmitt}, J.~H.~M.~M., \& {Wolk}, S.~J. 2013, \apj, 773,
  62

\bibitem[{{Redfield} {et~al.}(2008){Redfield}, {Endl}, {Cochran}, \&
  {Koesterke}}]{Redfield2008}
{Redfield}, S., {Endl}, M., {Cochran}, W.~D., \& {Koesterke}, L. 2008, \apjl,
  673, L87

\bibitem[{{Rossiter}(1924)}]{Rossiter1924}
{Rossiter}, R.~A. 1924, \apj, 60, 15

\bibitem[{{Sanchis-Ojeda} \& {Winn}(2011)}]{SanchisOjeda2011}
{Sanchis-Ojeda}, R. \& {Winn}, J.~N. 2011, \apj, 743, 61

\bibitem[{{Schlawin} {et~al.}(2010){Schlawin}, {Agol}, {Walkowicz}, {Covey}, \&
  {Lloyd}}]{Schlawin2010}
{Schlawin}, E., {Agol}, E., {Walkowicz}, L.~M., {Covey}, K., \& {Lloyd}, J.~P.
  2010, \apjl, 722, L75

\bibitem[{{Schmitt} {et~al.}(1995){Schmitt}, {Fleming}, \&
  {Giampapa}}]{Schmitt1995}
{Schmitt}, J.~H.~M.~M., {Fleming}, T.~A., \& {Giampapa}, M.~S. 1995, \apj, 450,
  392

\bibitem[{{Seager} \& {Sasselov}(2000)}]{Seager2000}
{Seager}, S. \& {Sasselov}, D.~D. 2000, \apj, 537, 916

\bibitem[{{Sing} {et~al.}(2012){Sing}, {Huitson}, {Lopez-Morales}, {Pont},
  {D{\'e}sert}, {Ehrenreich}, {Wilson}, {Ballester}, {Fortney}, {Lecavelier des
  Etangs}, \& {Vidal-Madjar}}]{Sing2012}
{Sing}, D.~K., {Huitson}, C.~M., {Lopez-Morales}, M., {et~al.} 2012, \mnras,
  426, 1663

\bibitem[{{Sing} {et~al.}(2011){Sing}, {Pont}, {Aigrain}, {Charbonneau},
  {D{\'e}sert}, {Gibson}, {Gilliland}, {Hayek}, {Henry}, {Knutson}, {Lecavelier
  Des Etangs}, {Mazeh}, \& {Shporer}}]{Sing2011}
{Sing}, D.~K., {Pont}, F., {Aigrain}, S., {et~al.} 2011, \mnras, 416, 1443

\bibitem[{{Sing} {et~al.}(2008){Sing}, {Vidal-Madjar}, {D{\'e}sert},
  {Lecavelier des Etangs}, \& {Ballester}}]{Sing2008LD}
{Sing}, D.~K., {Vidal-Madjar}, A., {D{\'e}sert}, J.-M., {Lecavelier des
  Etangs}, A., \& {Ballester}, G. 2008, \apj, 686, 658

\bibitem[{{Smette} {et~al.}(2015){Smette}, {Sana}, {Noll}, {Horst}, {Kausch},
  {Kimeswenger}, {Barden}, {Szyszka}, {Jones}, {Gallenne}, {Vinther},
  {Ballester}, \& {Taylor}}]{Smette2015}
{Smette}, A., {Sana}, H., {Noll}, S., {et~al.} 2015, ArXiv e-prints
  [\eprint[arXiv]{1501.07239}]

\bibitem[{{Snellen} {et~al.}(2008){Snellen}, {Albrecht}, {de Mooij}, \& {Le
  Poole}}]{Snellen2008}
{Snellen}, I.~A.~G., {Albrecht}, S., {de Mooij}, E.~J.~W., \& {Le Poole}, R.~S.
  2008, \aap, 487, 357

\bibitem[{{Snellen} {et~al.}(2010){Snellen}, {de Kok}, {de Mooij}, \&
  {Albrecht}}]{Snellen2010}
{Snellen}, I.~A.~G., {de Kok}, R.~J., {de Mooij}, E.~J.~W., \& {Albrecht}, S.
  2010, \nat, 465, 1049

\bibitem[{{Torres} {et~al.}(2008){Torres}, {Winn}, \& {Holman}}]{Torres2008}
{Torres}, G., {Winn}, J.~N., \& {Holman}, M.~J. 2008, \apj, 677, 1324

\bibitem[{{Triaud} {et~al.}(2009){Triaud}, {Queloz}, {Bouchy}, {Moutou},
  {Collier Cameron}, {Claret}, {Barge}, {Benz}, {Deleuil}, {Guillot},
  {H{\'e}brard}, {Lecavelier Des {\'E}tangs}, {Lovis}, {Mayor}, {Pepe}, \&
  {Udry}}]{Triaud2009}
{Triaud}, A.~H.~M.~J., {Queloz}, D., {Bouchy}, F., {et~al.} 2009, \aap, 506,
  377

\bibitem[{{Vidal-Madjar} {et~al.}(2003){Vidal-Madjar}, {Lecavelier des Etangs},
  {D{\'e}sert}, {Ballester}, {Ferlet}, {H{\'e}brard}, \& {Mayor}}]{Vidal2003}
{Vidal-Madjar}, A., {Lecavelier des Etangs}, A., {D{\'e}sert}, J.-M., {et~al.}
  2003, \nat, 422, 143

\bibitem[{{Vogt} {et~al.}(1987){Vogt}, {Penrod}, \& {Hatzes}}]{Vogt1987}
{Vogt}, S.~S., {Penrod}, G.~D., \& {Hatzes}, A.~P. 1987, \apj, 321, 496

\bibitem[{{Watson} {et~al.}(1981){Watson}, {Donahue}, \& {Walker}}]{Watson1981}
{Watson}, A.~J., {Donahue}, T.~M., \& {Walker}, J.~C.~G. 1981, \icarus, 48, 150

\bibitem[{{Winn} {et~al.}(2006){Winn}, {Johnson}, {Marcy}, {Butler}, {Vogt},
  {Henry}, {Roussanova}, {Holman}, {Enya}, {Narita}, {Suto}, \&
  {Turner}}]{Winn2006}
{Winn}, J.~N., {Johnson}, J.~A., {Marcy}, G.~W., {et~al.} 2006, \apjl, 653, L69

\bibitem[{{Wood} {et~al.}(2011){Wood}, {Maxted}, {Smalley}, \&
  {Iro}}]{Wood2011}
{Wood}, P.~L., {Maxted}, P.~F.~L., {Smalley}, B., \& {Iro}, N. 2011, \mnras,
  412, 2376

\bibitem[{{Wright} {et~al.}(2004){Wright}, {Marcy}, {Butler}, \&
  {Vogt}}]{Wright2004}
{Wright}, J.~T., {Marcy}, G.~W., {Butler}, R.~P., \& {Vogt}, S.~S. 2004, \apjs,
  152, 261

\bibitem[{{Wyttenbach} {et~al.}(2015){Wyttenbach}, {Ehrenreich}, {Lovis},
  {Udry}, \& {Pepe}}]{Wyttenbach2015}
{Wyttenbach}, A., {Ehrenreich}, D., {Lovis}, C., {Udry}, S., \& {Pepe}, F.
  2015, ArXiv e-prints [\eprint[arXiv]{1503.05581}]

\bibitem[{{Yan} {et~al.}(2015){Yan}, {Fosbury}, {Petr-Gotzens}, {Zhao}, \&
  {Pall{\'e}}}]{Yan2015}
{Yan}, F., {Fosbury}, R.~A.~E., {Petr-Gotzens}, M.~G., {Zhao}, G., \&
  {Pall{\'e}}, E. 2015, \aap, 574, A94

\bibitem[{{Zhou} \& {Bayliss}(2012)}]{Zhou2012}
{Zhou}, G. \& {Bayliss}, D.~D.~R. 2012, \mnras, 426, 2483

\end{thebibliography}

\end{document}